\documentclass[preprint,tightenlines,notitlepage,secnumarabic,amssymb,nobibnotes,nofootinbib,aps,prfluids,superscriptaddress,floatfix]{revtex4-2}
\usepackage{graphicx}%
\usepackage{dcolumn}%
\usepackage{bm}%

\usepackage[utf8]{inputenc}
\usepackage[T1]{fontenc}
\usepackage{booktabs,array,mathptmx,tabularx,lipsum,amsmath,multirow}
\usepackage{siunitx, xcolor}
\usepackage[version=4]{mhchem}
\usepackage{orcidlink}
\usepackage{hyperref}
\usepackage{cleveref}
\crefname{figure}{Fig.}{Figs.}
\Crefname{figure}{Figure}{Figures}
\crefname{equation}{Eq.}{Eqs.}
\Crefname{equation}{Equation}{Equations}
\crefname{chapter}{Chap.}{Chaps.}
\Crefname{chapter}{Chapter}{Chapters}
\crefname{table}{Tab.}{Tabs.}
\Crefname{table}{Table}{Tables}
\crefname{section}{Sec.}{Secs.}
\Crefname{section}{Section}{Sections}
\crefname{appendix}{App.}{Apps.}
\Crefname{appendix}{Appendix}{Appendices}
\hypersetup{hidelinks,pdftitle={Flow reorganization and transport enhancement in two-dimensional horizontal convection near a density extremum},pdfauthor={Zhiyang Cai, Shengqi Zhang, Kaizhen Shi, Zhouxin Jiang, Shijun Liao}}

\begin{document}

\title{Flow reorganization and transport enhancement in two-dimensional horizontal convection near a density extremum}

\author{Zhiyang Cai}
\affiliation{School of Ocean and Civil Engineering, Shanghai Jiao Tong University, Shanghai 200240, China}
\affiliation{Zhejiang Key Laboratory of Industrial Intelligence and Digital Twin, Eastern Institute of Technology, Ningbo 315200, China}
\affiliation{Ningbo Key Laboratory of Advanced Manufacturing Simulation, Eastern Institute of Technology, Ningbo 315200, China}

\author{Shengqi Zhang\orcidlink{0000-0001-8273-7484}}
\email[Contact author: ]{szhang@eitech.edu.cn}
\affiliation{Zhejiang Key Laboratory of Industrial Intelligence and Digital Twin, Eastern Institute of Technology, Ningbo 315200, China}
\affiliation{Ningbo Key Laboratory of Advanced Manufacturing Simulation, Eastern Institute of Technology, Ningbo 315200, China}

\author{Kaizhen Shi\orcidlink{0009-0005-7729-3125}}
\affiliation{School of Ocean and Civil Engineering, Shanghai Jiao Tong University, Shanghai 200240, China}
\affiliation{Zhejiang Key Laboratory of Industrial Intelligence and Digital Twin, Eastern Institute of Technology, Ningbo 315200, China}
\affiliation{Ningbo Key Laboratory of Advanced Manufacturing Simulation, Eastern Institute of Technology, Ningbo 315200, China}
\author{Zhouxin Jiang}
\affiliation{Zhejiang Key Laboratory of Industrial Intelligence and Digital Twin, Eastern Institute of Technology, Ningbo 315200, China}
\affiliation{Ningbo Key Laboratory of Advanced Manufacturing Simulation, Eastern Institute of Technology, Ningbo 315200, China}
\affiliation{Department of Aeronautical and Aviation Engineering, The Hong Kong Polytechnic University, Hong Kong SAR, China}
\author{Shijun Liao\orcidlink{0000-0002-2372-9502}}
\affiliation{School of Ocean and Civil Engineering, Shanghai Jiao Tong University, Shanghai 200240, China}

\begin{abstract}
Horizontal convection (HC) serves as a canonical model for geophysical and industrial flows driven by differential heating along a surface. While the classical Oberbeck--Boussinesq (OB) approximation is well established, the impact of a nonlinear equation of state, specifically the density extremum of water near $4^\circ\mathrm{C}$, remains underexplored. Here, we investigate this effect using two-dimensional numerical simulations over the Rayleigh-number range $10^6 \le Ra \le 5\times 10^{10}$. We examine four configurations, contrasting extremum (EXT) and monotonic (MON) buoyancy boundary conditions with linear (LE) and nonlinear (NE) equations of state. Our results reveal that, owing to strong cabbeling, the EXT-NE case develops mixing plumes that drive a strong bicellular circulation. At $Pr=1$, as $Re$ increases, this bicellular state becomes unstable and transitions to a single-roll circulation. This reorganization manifests as transitional anomalies in the Reynolds-number ($Re$) scaling, while the emergence of full-depth plumes alters the heat-transport mechanism. Consequently, in contrast to the classical Rossby scaling ($Nu \sim Ra^{1/5}$) observed in the reference cases, the EXT-NE case exhibits enhanced heat-transport scaling ranging from $Nu \sim Ra^{1/4}$ to $Nu \sim Ra^{1/3}$. To interpret this behavior, we examine the total energy budget and identify an additional potential-energy transfer term, $\Phi_{i2}$, arising from the nonlinear equation of state. The scaling argument suggests that the magnitude of this contribution is controlled by the characteristic plume height ($\hat{z}$). Specifically, when the plumes penetrate the entire cavity depth ($\hat{z} \sim H$), as observed in the EXT-NE case, the global kinetic-energy dissipation is no longer described by the standard OB-HC energy closure alone. The resulting model captures the main trends of the numerical data and provides a possible energy-budget interpretation of the enhanced transport observed in this two-dimensional configuration.
\end{abstract}

\maketitle

\section{Introduction}
Horizontal convection (HC), a flow driven by differential heating and cooling imposed on a single horizontal surface \cite{hughes2008horizontal}, is a canonical model for investigating buoyancy-driven flows in geophysical and industrial systems. Over the past several decades, HC has been extensively studied owing to its ubiquity in nature, including atmospheric circulation, lake convection, the oceanic meridional overturning circulation (MOC) \cite{macdonald1996,ding2022effect}, and flows induced by the urban heat island effect \cite{noto2023stratified}. It also plays a pivotal role in industrial processes, particularly in glass melting furnaces \cite{chiu2008very}.

The dynamics of HC are governed by the Rayleigh ($Ra$) and Prandtl ($Pr$) numbers. The flow is characterized by a distinctive large-scale asymmetric circulation \citep{rossby1965thermal,hughes2007theoretical,Mullarney2004}. This structure typically manifests as a localized plume (often termed the endwall plume) at the destabilizing buoyancy source \cite{hughes2008horizontal}, coexisting with a stably stratified interior and a broad, slow return flow. In the laminar regime, this circulation is steady and driven by boundary-layer dynamics \citep{rossby1965thermal,yan2021thermal}. However, as $Ra$ increases, the system undergoes a series of stability transitions, with the endwall plume serving as the primary initiation site for instabilities \cite{gayen2014stability,tsai2016origin,yan2023transitional}. At sufficiently high $Ra$, the flow evolves into a turbulent state characterized by chaotic plumes, unsteady eddies, and the concurrent emergence of both convective and shear instabilities \citep{Mullarney2004, gayen2014stability, passaggia2024limiting}.

Beyond characterizing flow patterns, a central objective of HC research is to establish quantitative predictions for transport efficiency. The foundational work by \citet{rossby1965thermal} balanced viscous and buoyancy forces within the boundary layer to deduce the classical heat transfer scaling for laminar HC: $Nu\sim Ra^{1/5}Pr^{0}$. This 1/5 scaling regime has been verified by numerous subsequent studies \citep{wang2005experimental, Mullarney2004}. For the inertia-dominated regime at high $Ra$, \citet{hughes2007theoretical} utilized a theoretical plume model to predict that the dependence on $Ra$ remains $1/5$, while the dependence on $Pr$ shifts to $1/5$, a result later confirmed numerically \citep{gayen2014stability}. Rigorous analysis by \citet{siggers2004bounds} established that the heat transport in the `ultimate regime' is bounded by a 1/3 power law. Subsequently, \citet{rocha2020improved} derived improved bounds for the system and recovered this same ultimate scaling exponent. \citet{shishkina2016heat} extended the Grossmann--Lohse (GL) theory \citep{grossmann2000scaling,grossmann2001thermal,grossmann2002prandtl,grossmann2004fluctuations} to HC, establishing the Shishkina--Grossmann--Lohse (SGL) framework. This theory successfully unifies the observed scaling laws across different parameter regimes \citep{ramme2019transition,passaggia2024limiting,passaggia2024limiting2}, recovering both the classical Rossby scaling ($Nu \sim Ra^{1/5}$) and the scaling predicted for the ultimate regime. Consequently, the SGL theory provides a comprehensive theoretical basis for HC under the Oberbeck--Boussinesq (OB) approximation.

However, realistic thermal convection frequently involves complex boundary conditions and non-Oberbeck--Boussinesq (NOB) effects \citep{chilla2012new}. A paradigmatic example is the density extremum of freshwater at approximately $4^\circ\text{C}$. This density anomaly strongly influences mixing and circulation in cold and seasonally ice-covered lakes \citep{bouffard2019convection,kirillin2012physics}, and is also central to convective flows coupled to freezing and melting in fresh water \citep{du2024physics}. Motivated by these natural systems, some idealized configurations, including penetrative-convection and cooling-box models, have shown that the density maximum can partition the water column into convecting and stably stratified regions and can substantially modify heat transport \citep{couston2017dynamics,toppaladoddi2018penetrative,olsthoorn2021cooling,PhysRevFluids.9.113501}.

Horizontal density gradients also arise naturally in cold-water environments. In ice-covered lakes, the spatially nonuniform release of heat stored in bottom sediments, together with sloping bathymetry, can generate lateral thermal gradients, downslope gravity currents, and basin-scale circulation \citep{huttula2010modelling,rizk2014basin}. These examples motivate the use of bottom-heated HC as an idealized framework for investigating the dynamical consequences of lateral thermal forcing in a fluid with a density maximum.

This phenomenon also has important implications for scaling theory. Although the SGL theory successfully unifies the scaling regimes of OB horizontal convection, it does not account for a nonlinear equation of state. It therefore remains unclear whether its energetic closure and transport predictions remain valid when the thermal forcing spans a density extremum. Here, we address this gap through numerical simulations of horizontal convection across the density extremum, using different combinations of temperature boundary conditions and linear/nonlinear equations of state to separate the effects of (i) the nonlinear equation of state and (ii) the nonmonotonic bottom buoyancy forcing. We focus on how their coupling reorganizes the large-scale flow, introduces an additional energy-conversion pathway, and modifies heat-transport scaling. The present configuration is not intended to reproduce any specific lake or ice-covered water body, but rather to isolate the coupling between horizontal thermal forcing and a nonlinear buoyancy--temperature relation. Effects such as rotation, phase change, salinity, radiative and geothermal heating, and spatially varying heat-flux boundary conditions are therefore neglected.

Previous HC studies have considered nonlinear equations of state, but in settings distinct from the present density-extremum problem. \citet{nycander2010horizontal} generalized the Paparella--Young energetic bound for an oceanic nonlinear equation of state, while \citet{ilicak2012simulations} compared linear and nonlinear oceanic equations of state in two-dimensional HC. These studies did not address horizontal thermal forcing that spans the temperature of maximum density, for which $db/dT$ changes sign. The unresolved problem is therefore not simply whether the equation of state is nonlinear, but how density-extremum-crossing forcing reorganizes HC flow topology, modifies potential-energy conversion, and changes heat-transport scaling. Recent work has further examined cold-water horizontal and vertical convection with realistic temperature-dependent material properties, including the nonlinear density, viscosity, and thermal conductivity of freshwater \citep{noto2026effects}. In that study, the HC configuration was driven by a laterally varying temperature imposed at the upper boundary and focused on the validity of constant-property OB-type modeling in ice-bounded water bodies. This differs from the present problem, where a continuous bottom temperature distribution is prescribed to isolate how density-extremum-crossing horizontal forcing reorganizes HC transport.

The remainder of the paper is organized as follows: \cref{sec:method} outlines the physical settings and governing equations; \cref{sec:result} analyzes the numerical results and underlying mechanisms; and \cref{sec:conclusion} concludes the paper.
\section{Governing equations and configuration}\label{sec:method}

\begin{figure}[t]
	\centering
	\includegraphics[width=1\linewidth]{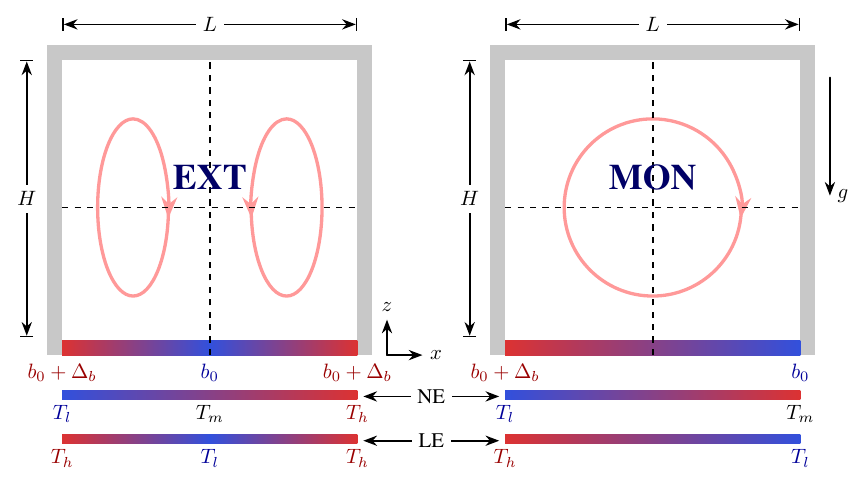}
\caption{
	Schematic of the computational domain and boundary conditions.
	The domain is a square enclosure with $L=H$.
	No-slip conditions are applied to all boundaries.
	A prescribed temperature profile (Dirichlet condition) is imposed on the bottom boundary,
	and the corresponding buoyancy profile follows from the equation of state,
	while the remaining walls are adiabatic.
	The parameters $b_0$, $T_l$, $T_m$, $T_h$, and $\Delta_b$ denote the reference
	buoyancy, cold boundary temperature, temperature of maximum density, hot boundary
	temperature, and the maximum buoyancy difference, respectively.
}
	\label{schematic}
\end{figure}

As illustrated in \cref{schematic}, the physical system consists of a square enclosure with an aspect ratio of $L:H=1:1$. A Cartesian coordinate system $\bm{x}=(x, z)$ is employed. The flow dynamics are governed by the incompressible Navier--Stokes and energy equations under the Boussinesq approximation, while incorporating a nonlinear equation of state \citep{wang2019penetrative}:
\begin{subequations}\label{eqn:NS_dim}
	\begin{gather}
		\nabla \cdot \bm{u} = 0, \label{mass} \\
		\frac{\partial\bm{u}}{\partial t} + \bm{u} \cdot \nabla\bm{u} =
		- \frac{1}{\rho_0}\nabla p + \nu \nabla^2\bm{u} + b(T)\bm{e}_z, \label{momentum} \\
		\frac{\partial T}{\partial t} + \bm{u} \cdot \nabla T = \kappa\nabla^2 T. \label{theta}
	\end{gather}
\end{subequations}
Here, $\bm{u}=(u,w)$, $p$, $T$, $b$, and $\bm{e}_z$ denote the velocity vector, pressure, temperature, buoyancy, and the vertical unit vector, respectively, while $\rho_0$ represents the reference density. The parameters $\nu$ and $\kappa$ are the kinematic viscosity and thermal diffusivity, respectively. The buoyancy is defined as $b=g(\rho_0-\rho)/\rho_0$, where $g$ is the gravitational acceleration. For the nonlinear equation of state (NE) cases, the density is modeled as $\rho=\rho_0[1-\beta_1(T-T_m)^2]$, where $T_m$ is the temperature of maximum density and $\beta_1$ characterizes the strength of the nonlinear density variation \citep{gebhart1977new,toppaladoddi2018penetrative}. For the linear equation of state (LE) cases, the density is instead given by $\rho=\rho_0[1-\beta_2(T-T_l)]$, where $T_l$ is the cold-boundary temperature and $\beta_2$ is the thermal expansion coefficient.

Classical horizontal convection typically adopts a linear equation of state (LE) and a linear temperature boundary condition \citep{rossby1965thermal}. In the present work, we instead consider HC with a linear temperature boundary condition and a nonlinear equation of state (NE), allowing the imposed temperature range to span a density extremum (EXT). The resulting EXT-NE configuration corresponds to strong cabbeling \citep{grace2021asymmetries,noto2026effects}. To isolate the underlying physical mechanisms, we introduce three controls. The MON-NE case retains the nonlinear equation of state but restricts the imposed temperature range to one side of the density maximum; it therefore represents weak cabbeling. The EXT-LE and MON-LE cases reproduce the corresponding extremal and monotonic bottom buoyancy profiles using a linear equation of state and thus do not exhibit cabbeling. These controls separate the influence of the nonlinear equation of state from that of the imposed bottom buoyancy distribution.

To facilitate a general analysis, the governing equations are non-dimensionalized using the free-fall velocity scale. Taking the enclosure length $L$ and the maximum buoyancy difference $\Delta_b$ as reference scales, the dimensionless variables are defined as:
\begin{equation}
	\tilde{\bm{x}} = \frac{\bm{x}}{L}, \quad \tilde{\bm{u}} = \frac{\bm{u}}{U_{f}}, \quad \tilde{t} = \frac{t}{L/U_{f}}, \quad \tilde{p} = \frac{p}{\rho_0 U_{f}^2},\quad \theta=\frac{T-T_l}{T_h-T_l}.
	\label{scaling_def}
\end{equation}
Here, the superscript $\tilde{\cdot}$ denotes dimensionless quantities. The characteristic free-fall velocity is given by $U_{f} = ({\Delta_b L})^{1/2}$; $T_h$ and $T_l$ are the imposed hot and cold boundary temperatures, respectively.
The dimensionless buoyancy term $\tilde{b}(\theta)$ is derived by normalizing the buoyancy difference by the characteristic buoyancy scale, $\Delta_b$. This scale is defined based on the maximum buoyancy difference across the bottom boundary:
\begin{itemize}
	\item[(i)] LE cases: For the standard Boussinesq approximation, the buoyancy scale is proportional to the temperature difference and is defined as $\Delta_b = g\beta_2(T_h - T_l)$. Consequently, the dimensionless buoyancy term simplifies to a linear relationship:
	\begin{equation}
		\tilde{b}(\theta) = \theta.
	\end{equation}

	\item[(ii)] NE cases: To capture the density extremum of water near $T_m=4^\circ\mathrm{C}$, the buoyancy scale is defined by a power law: $\Delta_b = g\beta_1\left( \max [ |T_h-T_m|,|T_l-T_m|] \right)^2$. In our simulations, where the dimensionless temperature of maximum density is set to $\theta_m=0.5$, this formulation leads to the following dimensionless relationship:
	\begin{equation}
		\tilde{b}(\theta) = \left| 2\theta - 1 \right|^2.
	\end{equation}
	For freshwater, $\Delta_b$ is not an independently adjustable forcing amplitude. Once the imposed temperature interval and the equation of state are specified, the maximum attainable density contrast over that interval is finite. Normalizing the buoyancy by $\Delta_b$ sets this finite contrast to unity, but does not imply that arbitrarily large density differences are physically realizable.
\end{itemize}
Substituting \cref{scaling_def} into \cref{eqn:NS_dim} yields the dimensionless governing equations:
\begin{subequations}\label{eqn:NS_nd}
	\begin{gather}
	\tilde{\nabla} \cdot \tilde{\bm{u}} = 0, \label{mass_nd} \\
	\frac{\partial \tilde{\bm{u}}}{\partial \tilde{t}} + (\tilde{\bm{u}} \cdot \tilde{\nabla}) \tilde{\bm{u}} =
	- \tilde{\nabla} \tilde{p} + \sqrt{\frac{Pr}{Ra}} \tilde{\nabla}^{2} \tilde{\bm{u}} + \tilde{b}(\theta)\bm{e}_z, \label{momentum_nd} \\
	\frac{\partial {\theta}}{\partial \tilde{t}} + (\tilde{\bm{u}} \cdot \tilde{\nabla}) {\theta} =
	\frac{1}{\sqrt{Ra Pr}} \tilde{\nabla}^{2} {\theta}. \label{theta_nd}
	\end{gather}
\end{subequations}
For fixed geometry, density-inversion parameter, and equation-of-state
exponent, the dimensionless dynamics are governed by the Prandtl number,
$Pr=\nu/\kappa$, and the buoyancy-based Rayleigh number,
$Ra=\Delta_b L^3/(\nu\kappa)$. Consequently, for fixed freshwater properties
and a fixed temperature interval, variation of $Ra$ should be interpreted
primarily as variation of the system size $L$, rather than as an unbounded
increase in $\Delta_b$. Because the present study focuses primarily
on the mechanisms responsible for the flow reorganization and the associated
change in heat transport, most simulations are performed at $Pr=1$. This
idealized value has also been adopted in previous studies of convection
involving a density maximum \citep{couston2017dynamics,toppaladoddi2018penetrative,ding2021coherent}.
Fixing $Pr$ limits the additional complexity associated with variations in
the relative thicknesses of the thermal and viscous boundary layers and
facilitates comparison with the classical Rossby-type scaling for horizontal
convection \citep{rossby1965thermal} and with the corresponding
$\mathrm{I}_l$ regime of the SGL framework
\citep{shishkina2016heat}. Nevertheless, $Pr=1$ is not quantitatively
representative of freshwater near its temperature of maximum density.
Additional simulations are therefore performed at $Pr=11.57$ to assess
whether the observed flow reorganization and heat-transport behavior persist
under conditions more representative of freshwater.

\Cref{eos} summarizes the specific combinations of equations of state and boundary conditions considered in this study. No-slip conditions are imposed on all walls, while adiabatic conditions are applied to the two vertical sidewalls and the top boundary. A prescribed temperature condition is imposed along the bottom boundary. In the primary EXT-NE configuration, the bottom temperature varies linearly in the horizontal direction, a profile that has been adopted in both laboratory experiments \citep{rossby1965thermal,wang2005experimental} and numerical simulations \citep{chiu2008very,ramme2019transition}. For all simulations, the initial conditions were set to $\bm{\tilde{u}}(\bm{\tilde{x}},0)={0}$ and $\theta(\bm{\tilde{x}},0)=0.5$ to shorten the initial transient while avoiding numerical instabilities associated with a strongly nonuniform initial temperature field \citep{rocha2020nusselt}. The dimensionless computational domain is a unit square, with $\tilde{x},\tilde{z}\in[0,1]$.

\begin{table*}[t]
	\centering
	\small
	\renewcommand{\arraystretch}{1}
	\setlength{\tabcolsep}{3pt}

\caption{Simulation cases characterizing the coupling between the buoyancy $\tilde{b}(\theta)$ and the bottom temperature distribution $\theta(\tilde{x})$ $(\tilde{x}\in[0,1])$. The dimensionless temperature of maximum density is $\theta_m=0.5$. EXT-NE and MON-NE correspond to strong and weak cabbeling, respectively, whereas the LE cases are non-cabbeling controls. The main dataset uses $Pr=1$, with additional EXT-NE and MON-NE simulations at $Pr=11.57$. The Rayleigh number ranges from $1\times 10^{6}$ to $5\times 10^{10}$. The pure conduction heat flux $Q_{\text{cond}}=\langle|\partial_z\theta_{Ra=0}|\rangle_A(\tilde z=0)$ for each case was calculated in Nek5000 on a $65\times65$ grid by disabling the buoyancy term.}
	\label{eos}

	\begin{tabular*}{\textwidth}
		{@{\extracolsep{\fill}} c c c c c c @{}}
		\toprule
		Case
		& Meaning
		& $\theta_{\mathrm{BC}}(\tilde{x})$
		& $\tilde{b}(\theta)$
		& $\tilde{b}_{\mathrm{BC}}(\tilde{x})$
		& $Q_{\mathrm{cond}}$
		\\
		\midrule

		EXT-NE
		& Extremal profile, nonlinear EOS (strong cabbeling)
		& $\tilde{x}$
		& $\left|2\theta-1\right|^2$
		& $\left|2\tilde{x}-1\right|^2$
		& $0.74$
		\\

		MON-NE
		& Monotonic profile, nonlinear EOS (weak cabbeling)
		& $\tilde{x}/2$
		& $\left|2\theta-1\right|^2$
		& $\left|1-\tilde{x}\right|^2$
		& $0.37$
		\\

		EXT-LE
		& Extremal profile, linear EOS (no cabbeling)
		& $\left|2\tilde{x}-1\right|^2$
		& $\theta$
		& $\left|2\tilde{x}-1\right|^2$
		& $1.65$
		\\

		MON-LE
		& Monotonic profile, linear EOS (no cabbeling)
		& $\left|1-\tilde{x}\right|^2$
		& $\theta$
		& $\left|1-\tilde{x}\right|^2$
		& $0.82$
		\\

		\bottomrule
	\end{tabular*}

\end{table*}

We characterize the system using two key dimensionless numbers: the Nusselt number ($Nu$), representing the heat transport efficiency, and the Reynolds number ($Re$), representing the flow intensity. The Nusselt number is defined as the ratio of the total heat flux to the purely conductive heat flux:
\begin{equation}\label{def:nu}
 Nu=\left.\dfrac{\langle{|\partial_z\theta|}\rangle_A}{\langle{|\partial_z\theta_{Ra=0}|}\rangle_A}\right|_{\tilde z=0},
\end{equation}
where the denominator corresponds to the conduction state ($Ra=0$) determined by the boundary conditions listed in \cref{eos}. Here, $\langle \cdot\rangle_A$ denotes the horizontal and temporal average. To characterize the strength of the large-scale circulation, the Reynolds number is defined based on the maximum horizontally averaged velocity:
\begin{equation}\label{def:re}
	 Re=\langle{|\tilde{u}|}\rangle_{A,\max}\sqrt{Ra/Pr}.
\end{equation}
For dimensional reference, \citet{moore1973nonlinear} represented freshwater over $0$--$8\,^{\circ}\mathrm{C}$ using the same quadratic equation of state, with $T_m=3.98\,^{\circ}\mathrm{C}$, $\rho_m=1000\,\mathrm{kg\,m^{-3}}$, and $\beta_1=8.0\times10^{-6}\,\mathrm{K^{-2}}$; this approximation is accurate to within about $4\%$ over that interval. Mapping the present $\theta_m=0.5$ to $T_l=0\,^{\circ}\mathrm{C}$ and $T_h\simeq8\,^{\circ}\mathrm{C}$ gives a maximum density contrast within the imposed interval of $\Delta\rho_{\max}\simeq\rho_m\beta_1(4\,\mathrm{K})^2\simeq0.128\,\mathrm{kg\,m^{-3}}$, corresponding to $\Delta_b\simeq1.26\times10^{-3}\,\mathrm{m\,s^{-2}}$. Using representative water properties $\nu\simeq1.567\times10^{-6}\,\mathrm{m^2\,s^{-1}}$ and $\kappa\simeq1.354\times10^{-7}\,\mathrm{m^2\,s^{-1}}$ ($Pr\simeq11.57$) \citep{hu2017flow}, the simulated range $10^6\le Ra\le5\times10^{10}$ corresponds to $L\simeq0.055$--$2.03\,\mathrm{m}$. Thus, under this freshwater mapping, increasing $Ra$ represents increasing the system size rather than exceeding the finite density contrast available over the prescribed temperature interval. This dimensional interpretation is illustrative and depends on the chosen physical temperature range.

\section{Results and discussion}\label{sec:result}

At the global level, the MON and LE reference configurations remain close to the classical horizontal-convection scalings, whereas EXT-NE exhibits a broader enhancement of heat transport and a localized anomaly in $Re$, as quantified later in \cref{Nu_re}. To interpret the enhanced heat transport and the localized anomaly in the Reynolds-number scaling observed in EXT-NE, the analysis is organized as follows:

\begin{itemize}

	\item
	We first examine the large-scale circulation and its evolution with $Ra$, focusing on the circulation transition and the enhanced flow intensity observed in the EXT-NE case.

	\item
	We then analyze the global energy budget to identify the energetic origin of the enhanced flow and to quantify the additional nonlinear potential-energy transfer term, $\Phi_{i2}$.

	\item
	Next, we examine the vertical distributions of the plumes and thermal dissipation to determine the characteristic height $\hat{z}$ entering the scaling estimate of $\Phi_{i2}$.

	\item
	Finally, we combine the resulting energy-budget closure with the GL framework to develop scaling estimates for $Nu$ and $Re$ and compare them with the numerical results.

\end{itemize}

\subsection{Flow organization}\label{sec:organization}

\begin{figure*}[htbp]
	\centering
	\includegraphics[width=1\linewidth]{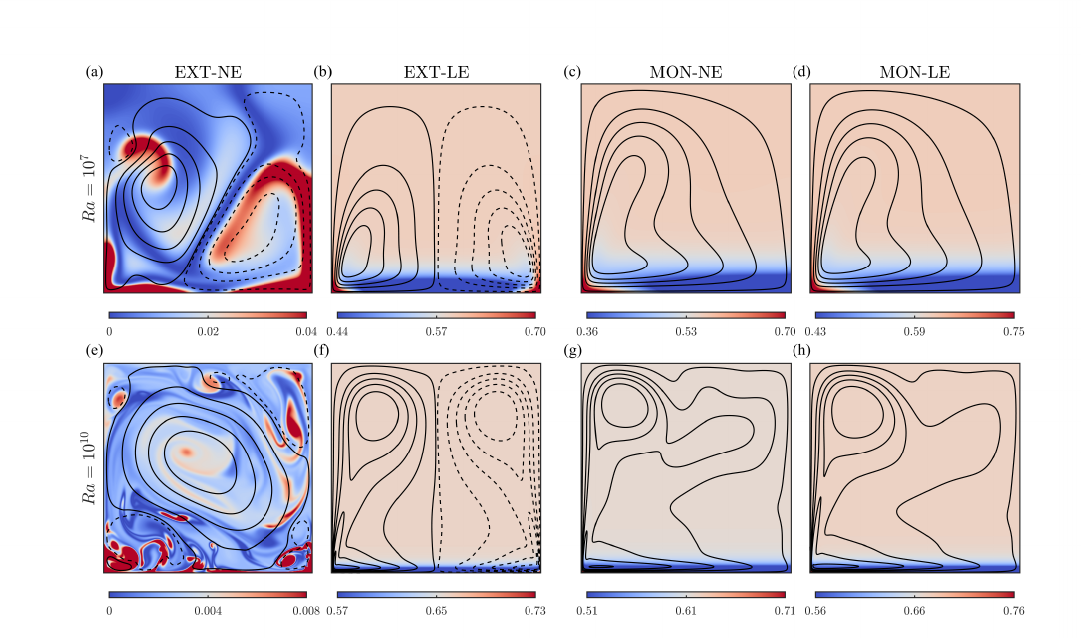}
\caption{Instantaneous dimensionless buoyancy fields (color contours) and streamlines for the EXT--NE, EXT--LE, MON--NE, and MON--LE cases (from left to right) at $Ra=10^7$ (top row) and $Ra=10^{10}$ (bottom row), with $Pr=1$. Solid and dashed streamlines indicate clockwise and counterclockwise circulation, respectively. To better reveal localized spatial variations, the color scale is adjusted independently for each panel. For each case, the representative snapshot was taken 2000 free-fall time units after the onset of statistical sampling.}
	\label{buoyancy}
\end{figure*}

In buoyancy-driven convection, the buoyancy field provides a direct indication of the dynamical mechanisms governing the flow. \Cref{buoyancy} shows instantaneous dimensionless buoyancy fields and the corresponding streamlines at $Ra=10^7$ and $10^{10}$. In classical horizontal convection, the upper part of the enclosure is occupied by a broad stably stratified region, which reduces the visual contrast of localized buoyancy anomalies when the full data range is used. To better reveal these structures, the color scale is adjusted independently for each panel.

The EXT-NE case exhibits a buoyancy distribution and flow structure that are markedly different from those of the other configurations. Positive buoyancy is generated near both ends of the bottom boundary. At relatively low $Ra$, a pronounced low-buoyancy region, with $\tilde{b}$ substantially smaller than that of the surrounding fluid, develops near the center of the domain. As $Ra$ increases to $10^{10}$, this central buoyancy anomaly disappears and is replaced by a more spatially homogeneous buoyancy distribution throughout the enclosure. This reorganization originates from the nonlinear equation of state, and its underlying mechanism will be discussed in detail below. By contrast, the other three reference cases exhibit broadly similar buoyancy distributions, characterized by a stably stratified configuration with larger buoyancy in the upper region and smaller buoyancy near the bottom. The resulting circulation is therefore relatively weak: appreciable buoyancy-driven upwelling is confined mainly to the corners of the bottom boundary layer, while the flow over the remainder of the domain is maintained primarily by a slow compensating return current \citep{hughes2008horizontal}.

\begin{figure*}[htbp]
	\centering
	\includegraphics[width=1\linewidth]{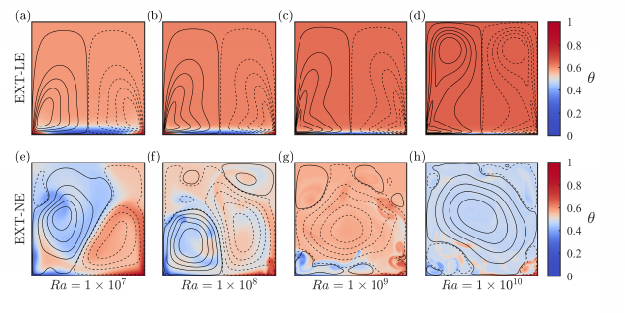}
\caption{Representative instantaneous temperature fields (color contours) and streamlines for the EXT cases at different Rayleigh numbers with $Pr=1$. Solid and dashed streamlines indicate clockwise and counterclockwise circulation, respectively. For each case, the snapshot was extracted 2000 free-fall time units after statistical sampling began.}

	\label{streamline_EXT}
\end{figure*}

\Cref{streamline_EXT} shows instantaneous snapshots of the temperature fields and corresponding streamline patterns for the EXT cases. In these EXT configurations, the difference in the equation of state leads to qualitatively different flow structures, despite identical buoyancy boundary conditions. The flow in the NE case is affected by strong cabbeling \citep{hanson2021stratified}. This effect, arising from thermal diffusion and mixing in a fluid with a nonlinear equation of state, generates plume-like downwellings, which we term `mixing plumes' (as shown in \cref{plume}). Unlike classical RBC, where plumes erupt from both top and bottom boundaries \cite{ahlers2009heat}, classical HC typically involves plume generation solely from the destabilizing boundary \citep{hughes2008horizontal}. In the EXT-NE case, these mixing plumes partially mimic the role of the downward-erupting plumes in RBC (\cref{plume}(a-d)), driving two large, counter-rotating vortices that span the full depth of the cavity. As $Ra$ increases, fluid carried by one branch of the initially symmetric circulation overshoots into the other half of the cavity, and the flow becomes time-periodic. With a further increase in $Ra$, cabbeling may amplify the instability over time, contributing to the reorganization of the double-vortex structure into a single large-scale roll reminiscent of the large-scale circulation (LSC) in classical RBC. This mixing-plume mechanism provides a plausible explanation for the distinct flow patterns observed.

For the EXT-LE case, a high density of streamlines is confined to the bottom and side walls, coinciding with the concentration of temperature gradients. This leaves the upper region of the domain stably stratified and the flow field within it largely inactive.

\begin{figure*}[t]
	\centering
	\includegraphics[width=1\linewidth]{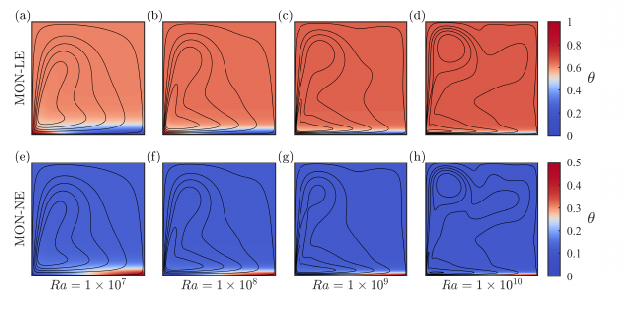}
\caption{Representative instantaneous temperature fields (color contours) and streamlines for the MON cases at different Rayleigh numbers with $Pr=1$. Solid and dashed streamlines indicate clockwise and counterclockwise circulation, respectively. For each case, the snapshot was extracted 2000 free-fall time units after statistical sampling began.}
	\label{streamline_MON}
\end{figure*}

Conversely, in the MON configurations (\cref{streamline_MON}), the flow structures for the LE and NE cases are qualitatively similar. Both cases exhibit a single, dominant circulation roll characteristic of classical HC. The main flow is concentrated in the lower region of the cavity, while a quiescent, stably stratified layer occupies the upper region \citep{hughes2008horizontal}.

The comparatively stable flows in the three reference cases (MON-NE, MON-LE, and EXT-LE) can be partly attributed to the buoyancy distribution intrinsic to horizontal convection \citep{hughes2008horizontal}. As the flow approaches a steady or statistically stationary state, high-buoyancy fluid accumulates in the upper part of the cavity, reducing the effective vertical buoyancy difference across the unstably stratified thermal boundary layer and thereby weakening its local buoyancy driving. Previous studies have shown that, at moderate-to-high $Pr$, unsteadiness in HC can first arise from a local Rayleigh--B\'enard-like instability of the thermally forced boundary layer \citep{tsai2016origin,gayen2014stability}. Therefore, the resulting reduction in the effective buoyancy difference, as illustrated in \cref{buoyancy,streamline_EXT,streamline_MON}, provides a physical explanation for the relative stability of the three reference cases.

The present results are qualitatively reminiscent of penetrative Rayleigh--B\'enard convection with a density maximum, for which variations in the density-inversion parameter can substantially reorganize the large-scale flow \citep{wang2019penetrative,wang2021universal}. The analogy should not, however, be extended too far because the forcing geometries are fundamentally different. In classical RBC, the lower hot and upper cold boundary layers generate ascending and descending plumes, respectively. In penetrative convection, the motions generated within the lower unstable layer penetrate into an overlying stably stratified region. Horizontal convection, by contrast, is driven by a buoyancy variation along a single horizontal boundary, for which only one part of the forcing is destabilizing \citep{hughes2008horizontal,reiter2020classical}. In the present bottom-forced monotonic configuration, this produces a dominant ascending plume, whereas the descending branch is primarily a compensating return flow. When the imposed temperature range spans the density maximum, mixing between fluids on opposite sides of the extremum creates anomalously dense fluid near the central part of the domain. The resulting descending ``mixing plume'' provides an active downward branch, strengthens the overturning circulation, and drives the observed reorganization of the large-scale flow.

\begin{figure*}
	\centering
	\includegraphics[width=1\linewidth]{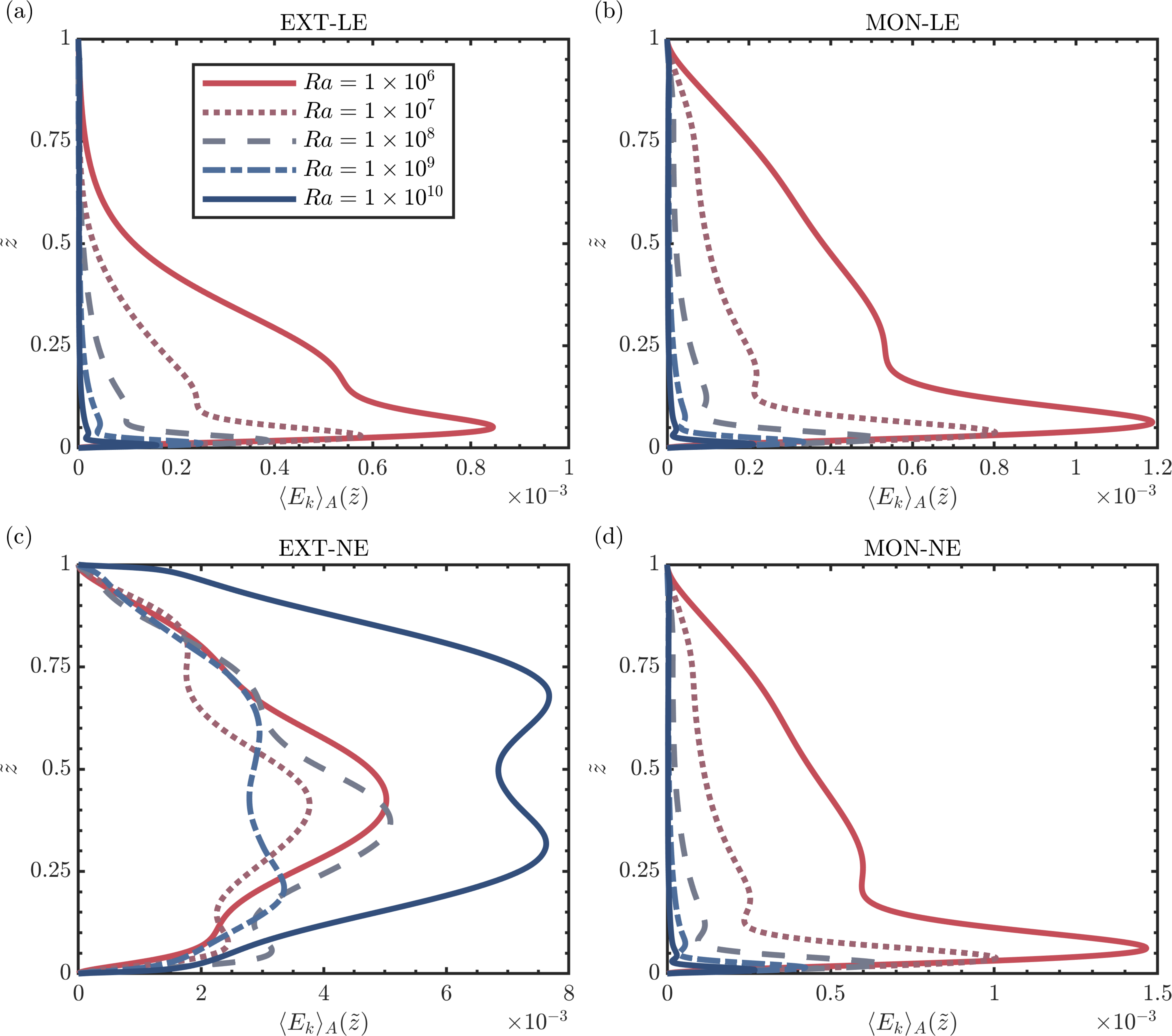}
\caption{Horizontally averaged vertical profiles of kinetic energy at different Rayleigh numbers with $Pr=1$.}
	\label{Ek}
\end{figure*}

To quantify these flow behaviors, we present the horizontally averaged vertical profiles of kinetic energy, defined as $\langle E_k\rangle_A(\tilde z)=\langle (\tilde{\bm{u}}\cdot \tilde{\bm{u}})\rangle_A/2$. As shown in \cref{Ek}(a,b,d), the kinetic energy generally diminishes with increasing $Ra$. Specifically, both the peak magnitude and the vertical extent of the active flow layer decrease, indicating a regime of partially penetrating convection \citep{wang2005experimental,chiu2008very}. In marked contrast, the EXT-NE case exhibits a full-depth distribution of kinetic energy that remains largely independent of $Ra$ when $Ra<1\times10^9$. Notably, at $Ra=1\times10^{10}$, a pronounced increase in kinetic energy is observed, accompanied by a vertically symmetric structure analogous to RBC (see \cref{streamline_EXT}(h)).

\Cref{streamline_EXT,Ek} indicate a clear transition in the global flow organization of the EXT-NE case. Given that different flow regimes (bicellular and unicellular) possess distinct heat transport efficiencies \citep{wei1994density}, we utilize Fourier mode decomposition to examine the flow energy composition \citep{hu2024flow}. Focusing on the four primary modes, we define the velocity basis functions as:
\begin{equation}
	\begin{split}
		u^{m,n} &= 2\sin(m\pi \tilde{x})\cos(n\pi \tilde{z}), \\
		w^{m,n} &= -2\cos(m\pi \tilde{x})\sin(n\pi \tilde{z}).
	\end{split}
\end{equation}
The kinetic energy $E_{m,n}(\tilde{t})$ contributed by the $(m,n)$ mode is given by:
\begin{equation}
	E_{m,n}(\tilde{t}) = \frac{1}{2}\left( [A_u^{m,n}(\tilde{t})]^2 + [A_w^{m,n}(\tilde{t})]^2 \right),
\end{equation}
where the time-dependent projection coefficients, $A_u^{m,n}(\tilde{t})$ and $A_w^{m,n}(\tilde{t})$, are defined by the spatial inner products: $A_u^{m,n}(\tilde{t})=\langle \tilde{u}(\tilde{t})u^{m,n}\rangle_{x,z}$ and $A_w^{m,n}(\tilde{t})=\langle \tilde{w}(\tilde{t})w^{m,n}\rangle_{x,z}$.

\begin{figure}[t]
	\centering
	\includegraphics[width=1\linewidth]{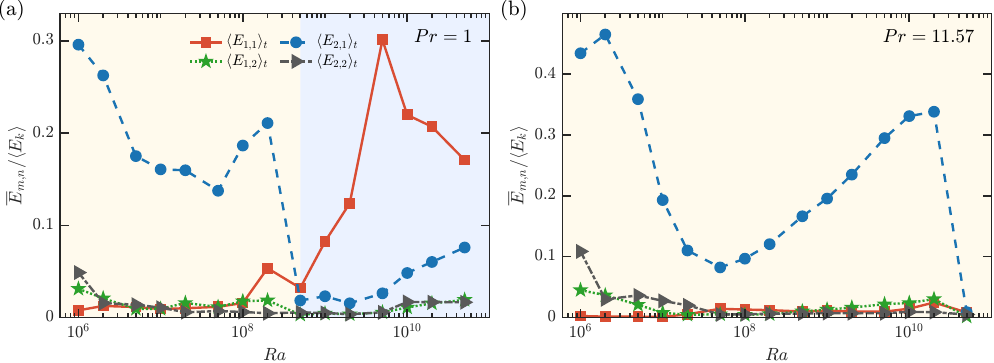}
\caption{Contributions of the \((1,1)\), \((1,2)\), \((2,1)\), and \((2,2)\) modes to the time-averaged kinetic energy at different Rayleigh numbers for the EXT-NE case: (a) \(Pr=1\); (b) \(Pr=11.57\).}
	\label{Emn}
\end{figure}
\Cref{Emn} illustrates the time-averaged relative contribution of each mode to the global kinetic energy, $\bar E_{m,n}/\langle E_{k}\rangle$, for the EXT-NE configuration, where $\bar\cdot $ denotes the time average and $\langle \cdot\rangle$ denotes the spatio-temporal average.
A significant transition in flow structure is observed at $Ra \approx 5 \times 10^8$ for $Pr=1$. Below this critical value, the flow is dominated by the (2,1) Fourier mode, manifesting as a bicellular structure with two vortices (see \cref{streamline_EXT}(e,f)). The interface between these counter-rotating vortices is marked by a central `mixing plume', driven by the density extremum. As $Ra$ exceeds this threshold, the flow evolves into a single-roll circulation dominated by the (1,1) mode (see \cref{streamline_EXT}(g,h)). This transition in flow organization accounts for the anomalous scaling behaviors presented in \cref{epu_ept,Nu_re}.

The bicellular-to-unicellular transition observed at \(Pr=1\) is not detected at \(Pr=11.57\) over the presently explored range of \(Ra\), suggesting that the selection of the large-scale circulation depends on the Prandtl number. More specifically, this dependence may arise primarily through the influence of \(Pr\) on the Reynolds number. The transition from a bicellular to a unicellular circulation may be associated with a shear instability, while increasing \(Pr\) reduces the flow velocity and hence weakens the velocity shear. As shown in \cref{Emn,tab:Ra_dependence,tab:simulation_parameters}, the Reynolds number at the transition point for \(Pr=1\) (\(Re=1279.92\)) is close to that attained at \(Ra=5\times10^{10}\) for \(Pr=11.57\) (\(Re=1232.70\)), where the amplitudes of all four modes become relatively small. This similarity suggests that the \(Pr=11.57\) cases may also undergo a transition to a unicellular circulation as \(Ra\) is increased further.

The results presented above show that the circulation transition coincides with a localized anomaly in $Re$, while the EXT-NE cases exhibit a systematically stronger flow intensity than the reference cases. This behavior suggests the presence of an additional pathway for energy transfer associated with the nonlinear equation of state. We therefore examine the global energy budget in the following section to clarify the origin of the enhanced motion.

\subsection{Global energy budget}

\begin{figure*}[t]
	\centering
	\includegraphics[width=1\linewidth]{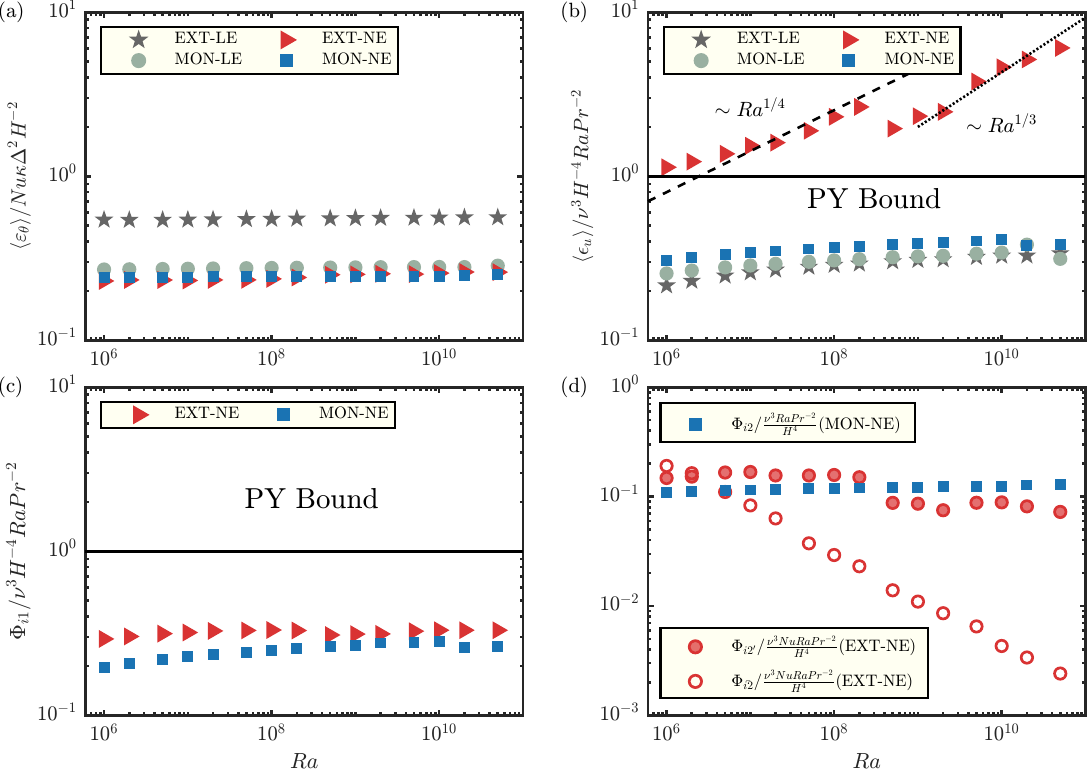}
	\caption{(a-d) Variations of the normalized quantities $\langle\varepsilon_\theta\rangle$, $\langle\varepsilon_u\rangle$, $\Phi_{i1}$, and $\Phi_{i2}$ with $Ra$, respectively, at $Pr=1$. The PY bound \citep{paparella2002horizontal} is given by \cref{ypbound}.}
	\label{epu_ept}
\end{figure*}

The kinetic ($\varepsilon_u$) and thermal ($\varepsilon_\theta$) dissipation rates are defined as:
\begin{equation}\label{varepsilon_ut}
	\varepsilon_u \equiv \nu (\partial_i u_j)^2,\qquad
	\varepsilon_\theta \equiv \kappa (\partial_i T)^2.
\end{equation}

For OB HC driven by the idealized step-function buoyancy profiles considered by \citet{shishkina2016heat}, the global dissipation rates are governed by the following exact balance and rigorous upper bound:
\begin{subequations}
\begin{gather}
		\langle\varepsilon_\theta\rangle = {\kappa\Delta^2Nu}/({2HL}), \label{varepsilon_t}\\
		\langle\varepsilon_u\rangle ={\langle{wb}\rangle} \leq {\nu^3 Ra Pr^{-2}}/{(2HL^3)}. \label{varepsilon_u}
\end{gather}
\end{subequations}
Here, $\Delta$ denotes the characteristic temperature difference. Regarding the scaling of global thermal dissipation, although the continuous buoyancy profiles employed in the present study introduce profile-dependent prefactors, the fundamental scaling dependence on the Nusselt number remains robust over a broad parameter range \citep{ding2021comparative}. This relationship is examined in \cref{epu_ept}(a). The data show good agreement with the theoretical scaling implied by \cref{varepsilon_t}, suggesting that the leading scaling exponent of the thermal dissipation is relatively insensitive to the boundary profile details considered here.

The scaling of the kinetic dissipation rate, $\langle\varepsilon_{u}\rangle$, shown in \cref{epu_ept}(b), reveals a more complex scenario. For the LE and MON configurations, the data adhere strictly to the theoretical upper bound given in \cref{varepsilon_u}. This classical limit is derived under the OB approximation, where the linear relation $b=\beta g(T-T_l)$ implies $\langle wb \rangle=\beta g\langle wT \rangle$. Utilizing the zero net heat flux property in HC, \citet{shishkina2016heat} derived $\langle\varepsilon_u\rangle\leq \nu^3RaPr^{-2}/(2HL^3)$, a result originally established by \citet{paparella2002horizontal} via energy conservation; we refer to this rigorous limit hereafter as the PY bound.

The EXT-NE case displays a dissipation behavior that is not captured by the standard OB-HC energy closure. This departure should not be interpreted as a violation of the PY bound; rather, it indicates that when the buoyancy is a nonlinear function of temperature, the OB energy budget lacks an additional conversion term. Building on the potential energy equation derived under the OB approximation \citep{winters1995available,siggia1994high}, we introduce a generalized potential energy equation to account for this contribution:

\begin{equation}
	\frac{\partial ({b}z)}{\partial t} + z  \bm{u} \cdot \mathbf{\nabla}  {{b}} = z{\kappa}\nabla^2T\frac{db}{dT},
\end{equation}
where the diffusion term accounts for the nonlinear equation of state. We then perform a decomposition analysis of the scaling law of $\langle wb \rangle$ by spatio-temporal averaging of the potential energy equation:
\begin{equation}\begin{split}
		\overbrace{\langle{w}{b}\rangle}^{{\scriptstyle -\Phi_{z}}}=
		\overbrace{-\frac{\kappa}{V}{\oint_{\partial{V}}  \left(z{\nabla}\bar b\right){\cdot} d\bm{S}}}^{ \scriptstyle \Phi_{b}}
		+\overbrace{\frac{\kappa}{V}{\oint_{\partial{V}} (\bar b\bm{e}_z){\cdot} d\bm{S}}}^{\scriptstyle \Phi_{i1}}
		+\overbrace{\kappa{\left\langle z\mathbf{\nabla}T{\cdot}{\nabla}\frac{db}{dT}\right\rangle}}
		^{\scriptstyle \Phi_{i2}},	\end{split}\label{Potential}
\end{equation}
where $\bar\cdot$ denotes the time average and $V$ is the volume of the cavity.

Combining \cref{varepsilon_u} and \cref{Potential}, we establish an exact relation between the global kinetic dissipation and potential energy transfer \citep{siggia1994high}:
\begin{equation}\label{epsUdecompose}
	\langle{\varepsilon_u}\rangle
	={-\Phi}_z ={\Phi}_{b}{+}~{\Phi}_{i1} {+}~{\Phi}_{i2},
\end{equation}
where $\Phi_{b}$, $\Phi_{i1}$, and $\Phi_{i2}$ represent potential energy transfers due to surface flux, irreversible heat conduction, and the NOB effect (nonlinear temperature-density relation), respectively.

$\Phi_b$ represents the net potential energy exchange through the boundaries. In HC, the net potential energy exchange across the boundaries vanishes, leading to $\Phi_b = 0$.

For a rectangular enclosure, $\Phi_{i1}$ simplifies to:
\begin{equation}
	{\Phi}_{i1}=\frac{\kappa}{H}
	{[\langle{b}\rangle_A(z=H)-\langle{b}\rangle_A(z=0)]}
	\leq{\frac{\nu^3}{H^{4}}RaPr^{-2}}.
	\label{ypbound}
\end{equation}
This term corresponds to the PY bound \citep{paparella2002horizontal}, which governs the total potential energy production in OB HC. Inspection of \cref{epu_ept}(c) reveals that this bound remains valid for the NE cases as well, but it is no longer the sole contributing term.

In OB HC, only thermal diffusion parallel to the gravity vector can alter the system's total potential energy, meaning that the potential energy transfer due to diffusion is solely attributable to the $\Phi_{i1}$ term.
However, when the NOB effects considered here are present, $db/dT$ is no longer constant, giving rise to the additional term $\Phi_{i2}$. Such nonlinear diffusive contributions are not unique to the present HC configuration, but arise more generally when buoyancy depends nonlinearly on temperature. Similar terms have been identified in energetic analyses of diffuse-interface Rayleigh--Taylor instability \citep{olsthoorn2019diffused} and radiatively heated ice-covered lakes \citep{winters2019energetics}.

Since $\Phi_{i2}$ originates from the thermal-diffusion term, its leading contribution is expected to arise from regions with large temperature gradients. As a scaling hypothesis, we therefore assume that $\Phi_{i2}$ is dominated by boundary-layer-like structures, including thermal boundary layers and detached plumes \citep{ahlers2009heat}. Following the Grossmann--Lohse (GL) approach of estimating dissipation from boundary-layer and bulk contributions \citep{grossmann2000scaling}, we estimate $\Phi_{i2}$ by substituting the corresponding characteristic scales, which yields
\begin{equation}\label{phii2}
	\Phi_{i2} \sim \kappa \, \hat{z} \, \frac{1}{\lambda_\theta^2} \Delta_b \frac{\lambda_\theta}{H}.
\end{equation}
Here, plumes are modeled as coherent structures that erupt from the boundary layer and thus inherit their characteristic geometric scales~\citep{grossmann2004fluctuations,castaing1989scaling,van2015plume}.
They are characterized by a length scale of the order of the system height, $H$, and a characteristic thickness, $\lambda_\theta$.
Consequently, the volume fraction occupied by these structures scales as $\lambda_\theta/H$. The characteristic buoyancy difference of these structures scales with the maximum buoyancy difference $\Delta_b$, and their characteristic vertical position is denoted by $\hat{z}$.
The scaling for the plume thickness is adopted from established theory for the thermal and kinetic boundary layers~\citep{grossmann2000scaling, PhysRevLett.114.114302}, which posits that $\lambda_\theta \sim H/Nu$ and $\lambda_u \sim H/Re^{1/2}$, respectively.

Thus, the contribution to $\Phi_{i2}$ can be broadly categorized as originating from two types of structures, depending on the dominant scale of $\hat z$: (i) boundary layers with a characteristic height scale comparable to the boundary layer thickness, $\hat {z}_b\sim\lambda_\theta$; and (ii) plumes with a characteristic height scale comparable to the system height, $\hat {z}_p\sim H$.

Substituting these distinct height scales into \cref{phii2}, we arrive at the following scaling relations:
\begin{equation}
	\Phi_{i2} \sim
	\begin{cases}
		\dfrac{\nu^3}{H^{4}}RaPr^{-2}, & \quad \text{for } \hat{z} \sim\hat {z}_b\sim \lambda_\theta, \\
		\\
		\dfrac{\nu^3}{H^{4}}{Nu}RaPr^{-2}, & \quad \text{for } \hat{z} \sim \hat {z}_p\sim H.
	\end{cases}
	\label{Phii2split}
\end{equation}
For the EXT-NE case, \cref{streamline_EXT,plume} show significant temperature gradients persisting throughout the cavity, suggesting that the dominant plume-related structures extend across the full cavity depth, implying $\hat{z}\sim H$ (see also \cref{epu_horizontal}(a)). Consequently, the plume-driven contribution of $\Phi_{i2}$ becomes dominant, establishing a non-negligible energy pathway.

In contrast, for the MON-NE case, the contribution of $\Phi_{i2}$ exhibits the same scaling as $\Phi_{i1}$ (\cref{epu_ept}(d)). This is consistent with \cref{streamline_MON}, where strong temperature gradients are mainly confined near the bottom boundary, implying $\hat{z}\sim\lambda_\theta$. A more detailed, quantitative characterization of the plume height distribution will be presented in \cref{sec:plume}.

To quantitatively isolate the role of these global-scale plumes, we distinguish their contribution from that of the background field, as the two are often intertwined in the bulk flow~\citep{grossmann2004fluctuations}. We therefore adapt the methodology proposed by \citet{ng2015vertical}, which decomposes the temperature field into its temporal average and fluctuating components ($T = \overline{T} + T^\prime$). In this framework, the mean contribution is defined as
\begin{equation*}
	{\Phi_{\bar{i2}}} = \left\langle \kappa z \nabla \overline{T} \cdot \nabla \left(\overline{ \frac{db}{dT} }\right) \right\rangle,
\end{equation*}
while the fluctuating component, $\Phi_{i2'}=\Phi_{i2}-{\Phi_{\bar{i2}}}$, serves as a proxy for the plume contribution.

Applying this decomposition to the EXT-NE case reveals that the fluctuating component, $\Phi_{i2'}$, scales as $\Phi_{i2'} \sim \nu^3 H^{-4} {Nu}\,{Ra}\,{Pr}^{-2}$, as shown in \cref{epu_ept}(d). In contrast, the contribution of the mean component, $\Phi_{\bar{i2}}$, decreases rapidly with increasing $Ra$. This confirms that the unique scaling in EXT-NE is dominated by the fluctuating field (associated with plumes). These findings provide strong quantitative support for our theory.

As outlined in \cref{epsUdecompose,ypbound,phii2,Phii2split}, we predict that the global kinetic dissipation rate in HC scales as
\begin{equation}
	\label{eq:epsu}\langle\varepsilon_u \rangle= {\Phi}_{b}+{\Phi}_{i1} + {\Phi}_{i2}\approx\left(\gamma_1 Nu + \gamma_2\right) \frac{\nu^3}{H^{4}}RaPr^{-2}.
\end{equation}
Here, $\gamma_1$ and $\gamma_2$ are prefactors determined primarily by the dominant height of the boundary-layer-like structures. Since $Nu$ increases with $Ra$ following a power law, the scaling behavior depends on whether $\gamma_1$ is significant. When $\gamma_1$ is non-negligible (corresponding to $\hat{z} \sim H$), the dissipation scales as ${\nu^3}{Nu}RaPr^{-2}/{H^{4}}$; otherwise, it scales as ${\nu^3}RaPr^{-2}/{H^{4}}$. These predictions also agree well with \cref{epu_ept}(b).

To simplify the subsequent derivation, we introduce an index $\eta$, which is determined by the essential boundary conditions, equation of state, and flow pattern. The unified scaling law for the kinetic dissipation can then be written as:
\begin{equation}\label{epsUnew}
	\langle\varepsilon_{u}\rangle \sim \frac{\nu^3}{H^{4}}Nu^{\eta}RaPr^{-2},
\end{equation}
where the index is given by:
\begin{equation}\label{epsUnewDiscuss}
	\eta=
	\begin{cases}
		1, & \text{EXT-NE},\\
		0, & \text{MON-NE \& LE}.
	\end{cases}
\end{equation}
As discussed above, $\eta=1$ corresponds to the EXT-NE case (where $\hat z\sim H$), whereas $\eta=0$ applies to the MON-NE case (where $\hat z\sim \lambda_\theta$) and the LE cases (for which $\langle\varepsilon_u\rangle=\Phi_{i1}$). The trends in \cref{epu_ept}(b) are consistent with this interpretation. It should be noted that $\eta$ is not a fitting parameter, but an index representing two asymptotic limiting balances for the dominant kinetic dissipation contribution. We therefore do not introduce an empirical interpolation function, because the present objective is to distinguish the two limiting flow states rather than to describe the detailed crossover between them.

In summary, our results reveal a strong connection between the global kinetic-energy dissipation and the characteristic plume height in the NE cases, where the additional $\Phi_{i2}$ term is present. We therefore examine the vertical extent and spatial distribution of the plume structures in greater detail in the following section to test this scaling estimate.

\subsection{Plume dynamics}\label{sec:plume}

Thermal plumes play a crucial role in the overall flow organization by clustering to form the LSC \citep{van2015plume}. In addition to this general dynamical function,
\cref{Phii2split} indicates that the plume height specifically affects the scaling behavior in the presence of a nonlinear equation of state. This motivates a detailed examination of plume dynamics.

\begin{figure*}[p]
	\centering
	\includegraphics[width=1\linewidth]{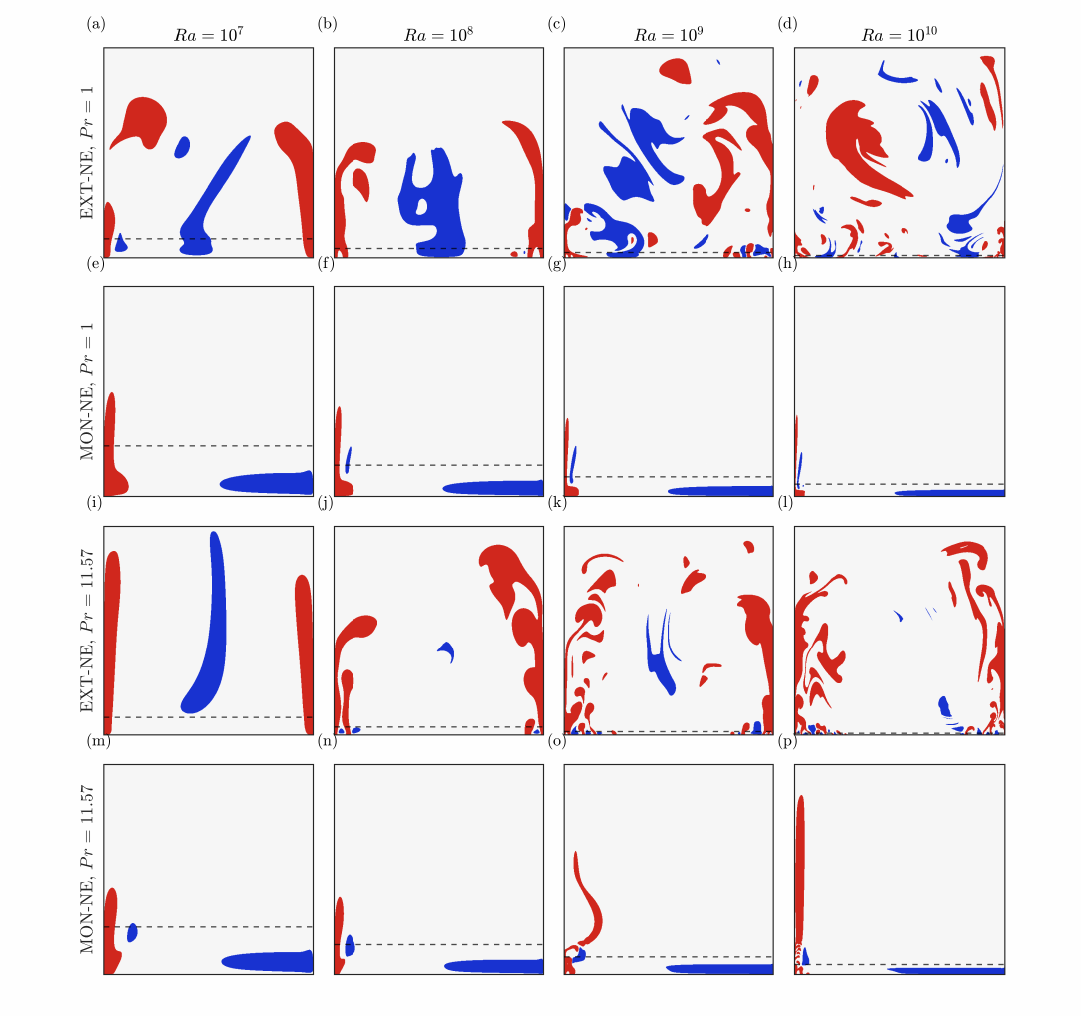}
	\caption{Representative instantaneous fields of identified plumes. From top to bottom, the rows show EXT-NE at $Pr=1$, MON-NE at $Pr=1$, EXT-NE at $Pr=11.57$, and MON-NE at $Pr=11.57$; from left to right, the columns correspond to $Ra=10^7$, $10^8$, $10^9$, and $10^{10}$. Red regions denote rising plumes with positive buoyancy fluctuation ($b^{\prime\prime}>0$), referred to as `endwall plumes'. Blue regions correspond to sinking plumes with negative buoyancy fluctuation ($b^{\prime\prime}<0$), termed `mixing plumes'. The dashed line indicates the approximate thermal-boundary-layer thickness, $\lambda_\theta=\Delta/\langle{|\partial_z\theta|}\rangle_{A}(0)$. For each case, the snapshot was extracted 2000 free-fall time units after statistical sampling began.}
	\label{plume}
\end{figure*}
For plume identification in RBC, criteria have been well established \citep{Zhang_Zhou_Sun_2017,van2015plume,emran2012conditional}. In this study, we employ a straightforward yet effective single-criterion approach developed by \citet{emran2012conditional}. To incorporate the nonlinear equation of state, the temperature fluctuation is replaced with the buoyancy fluctuation, defined as ${b}^{\prime\prime}({{\bm{x}}}, t)={b}({\bm{x}}, t)-\langle{{b}}\rangle_A({z})$. The threshold is defined as the product of a critical constant and the global kinetic dissipation ($\langle\varepsilon_{u}\rangle=\langle wb\rangle$). The plume-dominated subset $V_{pl}$ is defined such that
\begin{equation}
	V_{pl}=\{{\bm{x}}\in V\mid{w}{b}^{\prime\prime}>c \langle{\varepsilon_{u}}\rangle\}.
\label{plume_criterion}
\end{equation}
According to \citet{emran2012conditional}, the results are largely insensitive to the choice of the threshold; hence a representative value is sufficient. We therefore set the critical constant to $c=1.2$. This criterion means that any fluid region where the local fluctuating buoyancy flux is sufficiently large is classified as part of a plume.

\begin{figure*}[t]
	\centering
	\includegraphics[width=1\linewidth]{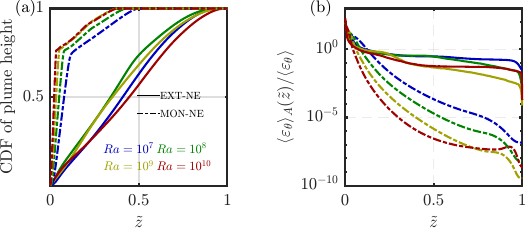}
\caption{(a) The cumulative distribution function (CDF) of the vertical coordinates of the identified plumes, $V_{\mathrm{pl}}$. (b) Vertical profiles of thermal dissipation normalized by the global mean. All results are obtained at $Pr=1$.}
	\label{epu_horizontal}
\end{figure*}

\Cref{plume} shows the plume distributions in the NE cases. In the EXT-NE cases, the plumes extend well beyond the thermal boundary layer. The ``mixing plumes'' are localized horizontally near the central region associated with the density extremum, whereas the ``endwall plumes'' rise along the lateral walls and may extend over the full cavity height. These full-depth plume structures help explain the distinct flow topology of the EXT-NE case discussed in \cref{sec:organization}. By contrast, in the MON-NE case at $Pr=1$, the characteristic plume height, $\hat{z}$, scales with the thermal-boundary-layer thickness, $\lambda_\theta$. Although the endwall plumes above the unstable source may extend slightly beyond the boundary layer, their vertical extent follows the boundary-layer scaling, such that $\hat{z}\sim\lambda_\theta$. This qualitative distinction is further supported by the cumulative distribution functions (CDFs) of the plume heights shown in \cref{epu_horizontal}(a). In the EXT-NE case, the plumes are distributed relatively uniformly throughout the vertical domain, whereas in the MON-NE case at $Pr=1$, they remain confined to a vertical extent that decreases with increasing $Ra$. For the MON-NE case at $Pr=11.57$, the endwall plumes penetrate to greater heights at large $Ra$, suggesting that the larger $Pr$ facilitates plume emission. Nevertheless, these plumes remain localized near the end wall and occupy only a limited fraction of the domain. Their influence on the global transport is therefore relatively weak, consistent with the $Nu$ and $Re$ results shown in \cref{Nu_re}.

Based on \cref{Potential}, $\Phi_{i2}$ can be expressed as $\Phi_{i2}= \kappa \left\langle z\, \nabla T \cdot \nabla\!\left({db}/{dT}\right) \right\rangle=\langle z\varepsilon_{\theta}(d^2b/dT^2)\rangle$, which highlights the correlation between $\Phi_{i2}$ and $\varepsilon_\theta$. Previous studies have demonstrated that high-amplitude events of thermal dissipation ($\varepsilon_\theta$) are co-located with plumes \citep{emran2012conditional, shishkina2006analysis}. Consequently, the $\Phi_{i2}$ distribution is also intrinsically linked to plume dynamics. This linkage is supported by the vertical distributions of $\varepsilon_{\theta}$ (normalized by $\langle\varepsilon_{\theta}\rangle$), which are presented in \cref{epu_horizontal}(b). As is evident from the figure, the ${\varepsilon_\theta}$ profiles of the MON-NE case decay more rapidly with $\tilde{z}$ than those of the EXT-NE case, a difference we attribute to variations in plume dynamics. It is noteworthy that although these vertical distributions of~$\varepsilon_{\theta}$ differ markedly, their global values share the same scaling (\cref{varepsilon_t}).

The key mechanism is that the distinct plume heights generate different thermal dissipation profiles, which, when weighted by the coordinate~${z}$, result in different values of~$\Phi_{i2}$. This observation provides qualitative support for the scaling arguments in \cref{Phii2split}.
\subsection{Scalings for $Nu$ and $Re$}
\begin{figure*}[t]
	\centering
	\includegraphics[width=1\linewidth]{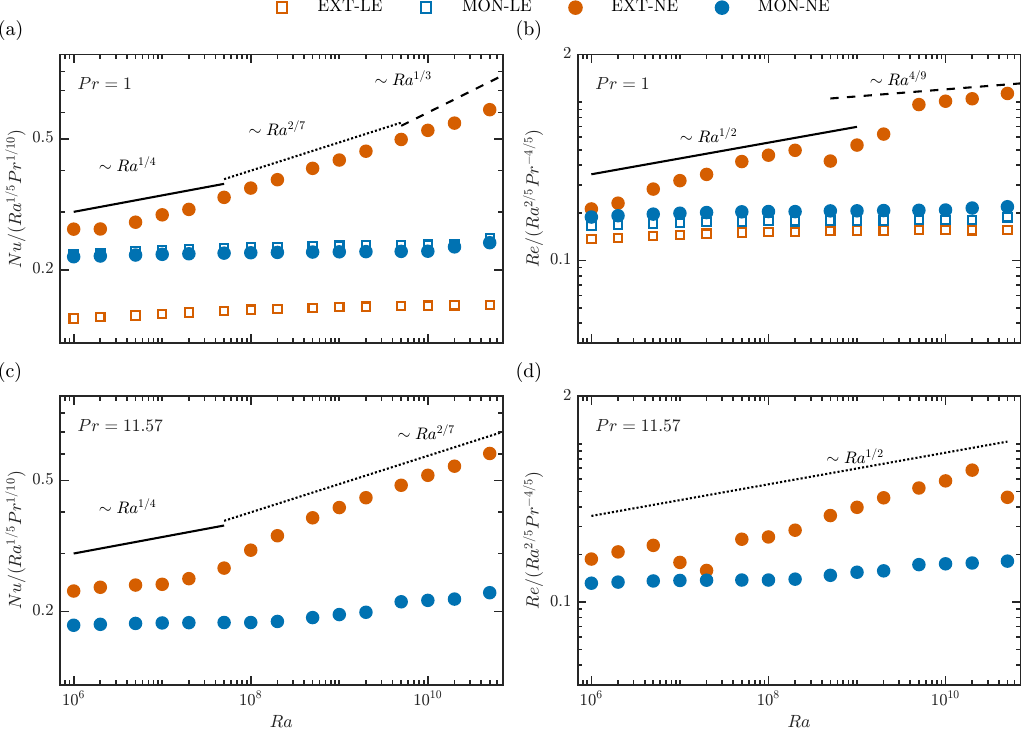}
\caption{Compensated Nusselt and Reynolds numbers as functions of $Ra$. (a,c) Compensated Nusselt number $Nu/(Ra^{1/5}Pr^{1/10})$ and (b,d) compensated Reynolds number $Re/(Ra^{2/5}Pr^{-4/5})$. The top row shows the four configurations at $Pr=1$, while the bottom row shows the EXT-NE and MON-NE cases at $Pr=11.57$. Solid, dotted, and dashed lines indicate the reference power-law scalings labeled in the corresponding panels.}
	\label{Nu_re}
\end{figure*}

The preceding analysis suggests that plume height controls the nonlinear potential-energy contribution to the global kinetic energy dissipation. We now combine the estimated dissipation closure (\cref{epsUnew}) with the GL framework~\citep{grossmann2000scaling,ahlers2009heat} to obtain scaling estimates for $Nu$ and $Re$ and compare them with the numerical results.

Following the regime classification of the GL theory, this study focuses on regimes $\text{I}_{l}$ and $\text{IV}_{u}$. In regime I, both the global kinetic and thermal energy dissipations are dominated by the boundary layers, whereas in regime IV, these dissipative processes are governed by the bulk flow. The subscripts `$l$' and `$u$' denote the relatively low- and high-$Pr$ regimes, respectively. To facilitate the scaling analysis, the globally averaged dissipation rates defined in \cref{varepsilon_ut} are decomposed into their respective boundary layer (BL) and bulk contributions. Specifically, the scaling laws governing regimes $\text{I}_{l}$ and $\text{IV}_{u}$ are expressed as follows:
\begin{align}\label{GLEpsU}
	\langle\varepsilon_u\rangle\sim
	\begin{cases}
			\langle\varepsilon_{u,\text{BL}}\rangle
			\sim \nu \left(\dfrac{\hat{u}}{\lambda_u}\right)^2 \dfrac{\lambda_u}{H}
			\sim \dfrac{\nu^{3}}{H^4}Re^{5/2}~(\text{I}_l),
			\\
			\\
			\langle\varepsilon_{u,\text{bulk}}\rangle
			\sim \dfrac{\hat{u}^3}{H}
			\sim \dfrac{\nu^{3}}{H^4}Re^{3}	~(\text{IV}_u).
		\end{cases}
\end{align}
Here, $\hat{u}$ denotes the wind velocity, which characterizes the $Re$ defined in \cref{def:re}.

In regime $\text{I}_{l}$, the advective-diffusive balance within the boundary layer ($u\partial_xT+w\partial_zT=\kappa\partial^2_zT$) implies:
\begin{equation}
	Nu\sim Re^{1/2} Pr^{1/2}.
	\label{theta_bl}
\end{equation}
Subsequently, by coupling the boundary layer scaling \cref{theta_bl} with the global kinetic energy dissipation closures derived in \cref{epsUnew,epsUnewDiscuss,GLEpsU}, we obtain the final scaling laws for the global transport coefficients:
\begin{equation}\begin{split}
		Nu \sim Ra^{1/(5-\eta)}Pr^{1/(10-2\eta)} , \quad Re \sim Ra^{2/(5-\eta)}Pr^{(\eta-4)/(5-\eta)},\label{scaling}
\end{split}\end{equation}
where $\eta=1$ for the EXT-NE case and $\eta=0$ for the reference cases (see \cref{epsUnewDiscuss}).

In regime $\text{IV}_{u}$, the thermal dissipation $\langle\varepsilon_\theta\rangle$ is dominated by the bulk flow, while the thermal boundary layer is nested within the kinetic boundary layer. This configuration implies the following scaling relation:
\begin{equation*}
	\langle \varepsilon_\theta \rangle \sim \hat{u} \dfrac{\lambda_\theta}{\lambda_u}\dfrac{\Delta^2}{H}
	\sim \kappa\dfrac{\Delta^2}{H^2}\dfrac{PrRe^{3/2}}{Nu}.
\end{equation*}
Coupling this relation with the established closures (\cref{varepsilon_t,epsUnew,epsUnewDiscuss,GLEpsU}) yields:
\begin{equation}
	Nu \sim Ra^{1/(4-\eta)}Pr^{0},\quad Re \sim Ra^{4/(12-3\eta)}Pr^{-2/3}.
\end{equation}

Classical OB-HC serves as a canonical model for studying thermal convection, with well-established scaling laws for both heat transport and flow intensity. The classical theory predicts scaling laws of $Nu \sim Ra^{1/5}$ and $Re \sim Ra^{2/5}$ for the heat transport and flow intensity, respectively \citep{rossby1965thermal,shishkina2016heat}, which have been validated over an extensive parameter range \citep{rossby1965thermal,hughes2008horizontal,shishkina2016heat,shishkina2016prandtl,passaggia2024limiting,passaggia2024limiting2}. To highlight deviations from this standard theory, we present the simulation results compensated by these classical exponents in \cref{Nu_re} (definitions of $Nu$ and $Re$ are provided in \cref{def:nu,def:re}). The compensated data reveal a striking contrast. The MON and LE configurations remain close to the classical OB-HC predictions, consistent with earlier evidence that the leading $Nu$ and $Re$ scalings are only weakly affected by changes in the imposed thermal-forcing profile. Nevertheless, such changes can still modify the scaling prefactors of global response quantities \citep{reiter2020classical,ding2021comparative,ramme2019transition}. By contrast, the EXT-NE case exhibits a pronounced departure from the classical scalings and follows distinct scaling laws. This behavior is consistent with our theoretical estimates, suggesting that the reconfiguration of global kinetic energy dissipation underlies the transition in transport scaling.

Within the GL framework, scaling relations transition between different regimes  \citep{ahlers2009heat}. As shown in \cref{Nu_re}(a), the EXT-NE case departs from the classical HC behavior: the effective scaling is approximately $Nu \sim Ra^{1/4}$ at lower $Ra$ and becomes steeper at higher $Ra$, showing a trend toward $Nu \sim Ra^{1/3}$. Rather than a sharp transition, the data exhibit a broad crossover reminiscent of that in RBC, where such crossovers often appear close to an effective $2/7$ scaling over an extended $Ra$ range \citep{wang2021regime,grossmann2000scaling,castaing1989scaling}. The Reynolds number shows a consistent trend, with the effective exponent evolving from slightly above $1/2$ to approximately $4/9$. The intermediate, non-asymptotic exponents are attributed to the gradual reorganization of the global flow (see \cref{Emn}); in particular, the drop in flow intensity near $Ra \approx 5\times 10^{8}$ coincides with the LSC transition identified in \cref{Emn}.

We further examine the EXT-NE and MON-NE cases at $Pr=11.57$, a value more representative of freshwater near its density maximum, as shown in \cref{Nu_re}(c,d). The results confirm that only the EXT-NE case exhibits a substantial enhancement relative to the classical predictions. Specifically, the effective scaling exponent of $Nu$ increases from approximately $1/4$ to $2/7$, while that of $Re$ approaches $1/2$. The effective scaling exponent of $Re$ likewise exhibits a local anomaly, consistent with the results presented in \cref{Emn}, indicating that the scaling of $Re$ is sensitive to the organization of the large-scale circulation. For EXT-NE at $Pr=11.57$, the local decrease in the compensated $Re$ around $Ra=10^7$ is associated with the transition from a steady laminar state to a time-dependent flow. We also observe a modest departure of the MON-NE results from the classical prediction at high $Ra$. This deviation may reflect the influence of the larger $Pr$ on the penetration of the endwall plume and the overall circulation.

\begin{figure*}[t]
	\centering
	\includegraphics[width=1\linewidth]{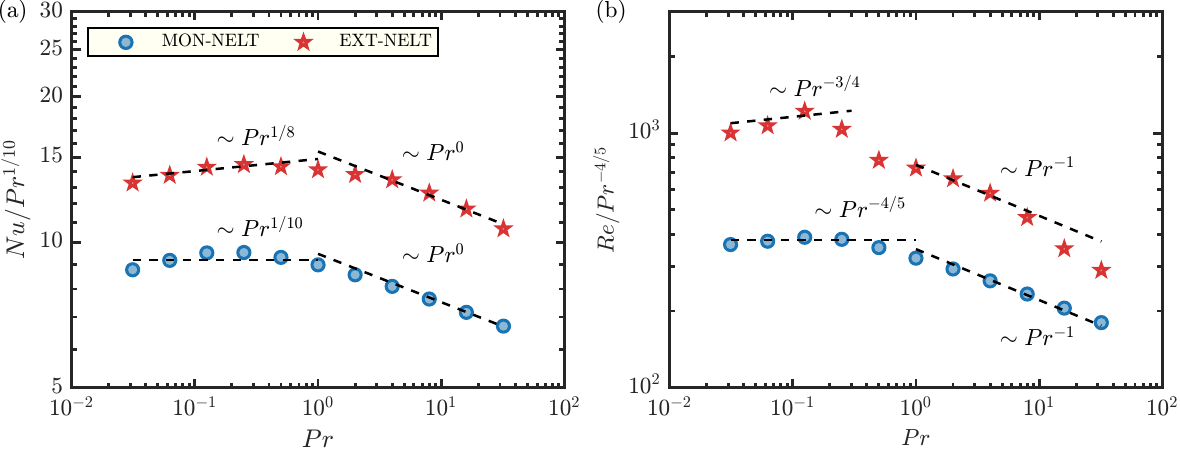}
\caption{
	(a) Compensated Nusselt number $Nu/Pr^{1/10}$ and (b) compensated Reynolds number $Re/Pr^{-4/5}$ as a function of $Pr$ for $Ra=1\times 10^{8}$.
}
	\label{Nu_re_pr}
\end{figure*}

To further validate the proposed scaling laws, we also examine the $Pr$-dependence of both $Nu$ and $Re$, with the results shown in \cref{Nu_re_pr}. However, the dependence on $Pr$ is not clearly distinguishable. This is consistent with the theoretical predictions in regime $\text{I}_{l}$, where the scaling exponents of $Pr$ for both $Nu$ and $Re$ differ only slightly (\cref{scaling}), making them difficult to resolve within the present parameter range. According to \citet{shishkina2016prandtl}, the transition between regimes $\text{I}_{l}$ and $\text{I}^*_{l}$ occurs near $Pr=1$. In our parameter range, the system spans these two regimes. In regime $\text{I}^*_{l}$, the expected scaling laws are $Nu \sim Pr^{0}$ and $Re \sim Pr^{-1}$, regardless of whether $\eta=0$ or $\eta=1$ (for a detailed derivation, see \citet{shishkina2016prandtl} together with \cref{epsUnew,epsUnewDiscuss}). The anomalous exponent observed in $Re$ for $Pr < 1$ can also be attributed to flow reorganization, as discussed in the $Ra$-dependence analysis. A more detailed investigation of additional $Pr$-dependent scaling regimes in horizontal convection is beyond the scope of the present study and will be addressed in future work.

In summary, these results demonstrate that HC with a temperature range spanning the density extremum undergoes a profound reconfiguration of its energetic pathways. The additional nonlinear conversion term increases the total kinetic dissipation beyond the contribution constrained by the classical PY bound for OB-HC \citep{paparella2002horizontal}. Consequently, this shift facilitates a transition to more efficient transport whose global scaling exponents markedly exceed the theoretical predictions for conventional HC.

\section{Conclusions}\label{sec:conclusion}

In this study, we have performed a two-dimensional numerical investigation of HC governed by a nonlinear equation of state near a density extremum. By comparing extremum and monotonic configurations with linear and nonlinear temperature-density relations, we isolate the role of the density extremum in modifying the large-scale flow organization, energy conversion, and heat transport. Within the explored parameter range, the simulations show that the EXT-NE configuration exhibits enhanced heat-transport scaling relative to the reference cases and that this enhancement remains robust at \(Pr=11.57\). These observations are interpreted using a global energy-budget analysis and the SGL framework.

The main findings are summarized as follows.

\begin{itemize}
	\item[(1)] Within the present set of configurations, the density extremum plays a central role in reorganizing the large-scale flow. In the EXT-NE case, the nonlinear temperature-density relation generates central `mixing plumes' associated with the density inversion. These plume-dominated structures penetrate into the bulk and promote a full-depth circulation. At $Pr=1$, the flow evolves from a bicellular structure toward a single-roll circulation as $Ra$ increases. This reorganization is associated with noticeable changes in $Re$ and the global kinetic energy dissipation.
	\item[(2)] The global energy budget contains an additional potential-energy transfer term associated with the nonlinear equation of state, denoted by $\Phi_{i2}$, which is absent from the standard OB horizontal-convection energy balance. We estimate $\Phi_{i2}$ by relating its magnitude to the effective vertical extent of the plume-dominated temperature-gradient structures. This argument leads to the following scaling for the kinetic-energy dissipation:
	\[
	\langle\varepsilon_{u}\rangle
	\sim
	\nu^3 H^{-4} Nu^{\eta} Ra Pr^{-2},
	\]
	where the index $\eta$ reflects whether the dominant plume activity remains confined near the boundary layer or extends over the cavity depth. When the effective plume height is of order $H$, as observed in the EXT-NE case, the kinetic energy dissipation is no longer described by the standard OB-HC energy closure alone.
	\item[(3)] The modified kinetic dissipation scaling provides a possible explanation for the transport trends observed in the simulations. The reference configurations remain close to the classical Rossby scaling for HC \citep{rossby1965thermal}, whereas the EXT-NE case exhibits a steeper effective $Ra$ dependence. Within the explored parameter range, the data are consistent with a crossover from an $\mathrm{I}_l$-type balance toward an $\mathrm{IV}_u$-type trend in the GL framework. This steeper scaling behavior in EXT-NE should not be regarded as a conventional SGL regime transition; rather, it reflects a modification of the global energy budget through the additional potential energy transfer associated with the nonlinear equation of state. This mechanism alters the kinetic dissipation balance and thereby changes the apparent transport scaling in EXT-NE. By contrast, a nonlinear equation of state in penetrative RBC has been reported not to substantially modify global transport exponents \citep{wang2019penetrative}.
	\item[(4)] The additional simulations at $Pr=11.57$ indicate that the transport enhancement is not restricted to $Pr=1$. Although increasing $Pr$ reduces $Re$, modifies the boundary-layer structure, and delays the large-scale flow reorganization, the EXT-NE case continues to exhibit stronger heat transport than the corresponding MON-NE case. The persistence of full-depth mixing plumes and enhanced transport is consistent with the proposed nonlinear potential-energy-transfer interpretation.
\end{itemize}

The proposed energy-budget interpretation is supported by the present simulations, but its quantitative validation remains incomplete. The available $Pr$-dependent data are broadly compatible with this interpretation, but they do not provide a definitive test of the predicted $Pr$ scalings because the differences between the predicted $Pr$ exponents are small and the explored parameter range is limited. Further simulations over a broader range of $Pr$ are therefore needed. More importantly, the present study is restricted to two-dimensional HC in a square enclosure; whether the plume-induced flow reorganization and the associated energy-transfer mechanism persist in three-dimensional configurations remains an important direction for future work.

\section*{Acknowledgments}
This work was supported by the National Natural Science Foundation of China (Grant No. 12302284, U25A6006), the Ningbo Municipal Bureau of Education (Grant No. 2024A-149-G), and the Ningbo Municipal Bureau of Science and Technology (Grant Nos. 2023Z227 and 2023-DST-001).
\section*{Declaration of interests}
The authors report no conflict of interest.
\section*{CRediT authorship contribution statement}
\noindent\textbf{Zhiyang Cai:} Conceptualization, Data curation, Formal analysis, Investigation, Methodology, Software, Validation, Visualization, Writing -- original draft, Writing -- review \& editing.
\textbf{Shengqi Zhang:} Conceptualization, Formal analysis, Funding acquisition, Methodology, Project administration, Resources, Supervision, Writing -- review \& editing.
\textbf{Kaizhen Shi:} Writing -- review \& editing.
\textbf{Zhouxin Jiang:} Writing -- review \& editing.
\textbf{Shijun Liao:} Funding acquisition, Writing -- review \& editing.
\section*{Data Availability}
The data that support the findings of this study are available from the corresponding author upon reasonable request.

\clearpage
\appendix
\section{Numerical details and grid resolution}

We used the spectral-element solver Nek5000 \citep{nek5000-web-page} to perform the two-dimensional simulations. Nek5000 has been widely employed in studies of buoyancy-driven convection \citep{couston2022competition,leard2020coupled}. Within each spectral element, the velocity and temperature fields were represented by seventh-order polynomials defined on eight Gauss--Lobatto--Legendre points in each coordinate direction, whereas the pressure field was represented by fifth-order polynomials defined on six Gauss--Legendre points. This discretization corresponds to the standard \(P_N\)-\(P_{N-2}\) velocity-pressure formulation used in Nek5000 \citep{deville2002high,nek5000-web-page}. Time integration was performed using a second-order semi-implicit scheme: the temporal derivatives were approximated using a second-order backward differentiation formula, the nonlinear advection terms were evaluated by second-order explicit extrapolation from the two preceding time levels, and the viscous and thermal diffusion terms were treated implicitly. The time step was adjusted dynamically to maintain a target Courant--Friedrichs--Lewy number of \(0.5\).

To reduce aliasing errors associated with the nonlinear advection terms, a \(3/2\) over-integration rule was employed. Although the solution fields were represented on eight Gauss--Lobatto--Legendre points in each element direction, the nonlinear terms were evaluated on an enlarged quadrature grid containing 12 points in each direction and were subsequently projected back onto the solution grid. No explicit spectral filter, artificial-viscosity model, or subgrid-scale model was employed. The numerical dissipation was therefore limited to that inherent in the spatial and temporal discretizations and the iterative solution procedures. The Helmholtz systems for the velocity and temperature fields were solved with a residual tolerance of \(10^{-8}\), while the pressure projection was performed with a divergence tolerance of \(10^{-6}\).

The spatial resolution was selected to resolve both the near-wall gradients and the smallest dynamically relevant dissipative structures. The detailed resolutions are summarized in \cref{tab:simulation_parameters}, where \(N_x\) and \(N_z\) denote the total numbers of grid points used for the velocity and temperature fields in the streamwise and wall-normal directions, respectively. The spectral elements were concentrated near the bottom boundary \citep{ramme2019transition}, where the velocity and temperature gradients and the associated viscous and thermal dissipation rates are largest \citep{couston2022competition}.

Grid-sensitivity tests were performed for all four configurations at $Pr=1$, as summarized in \cref{tab:simulation_parameters}. Relative changes are calculated as the absolute difference between the two grid results divided by the larger value. Further mesh refinement changes $Nu$ by less than $1\%$ in all tested cases. For the three reference configurations (MON-NE, EXT-LE, and MON-LE), the changes in both $Nu$ and $Re$ are below $0.5\%$. After the initial transients had decayed, steady cases were monitored for 5000 free-fall time units, whereas statistics for unsteady cases were accumulated over 2000 free-fall time units after the flow had reached a statistically stationary state.

\begin{table*}[!htbp]	\label{detail}
	\centering
	\caption{Simulation parameters, grid resolutions, Nusselt numbers $Nu$ and Reynolds numbers $Re$. The Prandtl number is fixed to $Pr=1$.}
	\label{tab:simulation_parameters}

	\footnotesize
	\setlength{\tabcolsep}{4pt}
	\renewcommand{\arraystretch}{1.0}

	\resizebox{\textwidth}{!}{%
		\begin{tabular}{c c c c c c c c c c}
			\toprule
			$Ra$ & $N_x\times N_z$
			& \multicolumn{4}{c}{$Nu$}
			& \multicolumn{4}{c}{$Re$} \\
			\cmidrule(lr){3-6}
			\cmidrule(lr){7-10}
			&
			& EXT-NE & MON-NE & EXT-LE & MON-LE
			& EXT-NE & MON-NE & EXT-LE & MON-LE \\
			\midrule
			$1 \times 10^{6}$ & 113$\times${113} & 4.21 & 3.48  & 2.26 & 3.53 &53.08 & 47.18 & 34.20 &41.47\\
			$1 \times 10^{6}$ & 141$\times${141} & 4.21 & 3.48  & 2.26 & 3.53 &53.05 & 47.18 & 34.29 & 41.47 \\
			$2 \times 10^{6}$ & 127$\times${127} & 4.85 & 4.01  & 2.62 & 4.09 &76.18 & 63.47 & 45.76 &55.61\\
			$5 \times 10^{6}$ & 127$\times${127} & 6.10 & 4.85  & 3.18 & 4.98 &134.92 & 93.71 & 68.17 &81.77\\
			$5 \times 10^{6}$ & 155$\times${155} & 6.14 & 4.85 & 3.18 & 4.98 &135.54 & 93.76 & 68.04 & 81.82 \\
			$1 \times 10^{7}$ & 141$\times${141} & 7.38 & 5.60  & 3.69 & 5.77 &201.28 & 125.29 & 91.41 &109.15\\
			$1 \times 10^{7}$ & 169$\times${169} & 7.38 & 5.60 & 3.69 & 5.77 &202.74 & 125.37 & 91.59 & 109.31 \\
			$2 \times 10^{7}$ & 169$\times${169} & 8.80 & 6.46  & 4.29 & 6.67 &290.91 & 166.94 & 122.76 &145.41\\
			$5 \times 10^{7}$ & 169$\times${169} & 11.50 & 7.79  & 5.21 & 8.08 &506.00 & 243.39 & 180.25 &211.36\\
			$5 \times 10^{7}$ & 204$\times${204} & 11.50 & 7.79 & 5.21 & 8.08 &503.10 & 243.59 & 180.24 & 211.69 \\
			$1 \times 10^{8}$ & 211$\times${211} & 14.10 & 8.98  & 6.03 & 9.32 &732.30 & 322.53 & 239.96 &280.22\\
			$1 \times 10^{8}$ & 253$\times${253} & 14.07 & 8.98 & 6.03 & 9.32 &730.92 & 322.66 & 239.78 & 279.85 \\
			$2 \times 10^{8}$ & 561$\times${561} & 17.17 & 10.33  & 6.97 & 10.75 &1037.75 & 425.11 & 317.28 &369.10\\
			$5 \times 10^{8}$ & 561$\times${561} & 22.33 & 12.45  & 8.43 & 12.97 &1279.92 & 619.73 & 463.64 &538.20\\
			$1 \times 10^{9}$ & 561$\times${561} & 27.18 & 14.32  & 9.73 & 14.95 &2131.30 & 818.59 & 614.20 &711.75\\
			$2 \times 10^{9}$ & 561$\times${561} & 33.21 & 16.47  & 11.22 & 17.21 &3300.23 & 1084.51 & 811.31 &938.11\\
			$5 \times 10^{9}$ & 561$\times${561} & 43.29 & 19.82  & 13.53 & 20.73 &7316.71 & 1570.49 & 1176.40 &1360.26\\
			$5 \times 10^{9}$ & 841$\times${841} & 43.46 & 19.82 & 13.53 & 20.73 &7264.93 & 1574.53 & 1176.34 & 1359.86 \\
			$1 \times 10^{10}$ & 841$\times${841} & 53.00 & 22.79  & 15.59 & 23.86 &10136.72 & 2080.83 & 1555.17 &1796.44\\
			$2 \times 10^{10}$ & 841$\times${841} & 64.04 & 26.98  & 17.94 & 27.44 &13862.71 & 2826.26 & 2041.78 &2376.60\\
			$5 \times 10^{10}$ & 841$\times${841} & 84.21 & 33.37  & 21.61 & 34.40 &21597.55 & 4154.33 & 2967.43 &3559.41\\
			$5 \times 10^{10}$ & 1009$\times${1009} & 83.62 & 33.52 & 21.59 & 34.49 &21426.04 & 4159.41 & 2974.59 & 3567.63\\
			\bottomrule
		\end{tabular}%
	}
\end{table*}

\begin{table*}[!htbp]
	\centering
	\caption{Simulation parameters, grid resolutions, Nusselt numbers ($Nu$), and Reynolds numbers ($Re$). The Prandtl number is fixed at $Pr = 11.57$.}
	\label{tab:Ra_dependence}

	\footnotesize
	\setlength{\tabcolsep}{3.5pt}
	\renewcommand{\arraystretch}{1.15}

	\resizebox{\textwidth}{!}{%
		\begin{tabular}{@{}c c c c c c @{\hspace{1.2em}} c c c c c c@{}}
			\toprule
			$Ra$ & $N_x \times N_z$
			& \multicolumn{2}{c}{$Nu$}
			& \multicolumn{2}{c}{$Re$}
			&
			$Ra$ & $N_x \times N_z$
			& \multicolumn{2}{c}{$Nu$}
			& \multicolumn{2}{c}{$Re$} \\
			\cmidrule(lr){3-4}
			\cmidrule(lr){5-6}
			\cmidrule(lr){9-10}
			\cmidrule(lr){11-12}
			&
			& EXT-NE & MON-NE
			& EXT-NE & MON-NE
			&
			&
			& EXT-NE & MON-NE
			& EXT-NE & MON-NE \\
			\midrule
$1\times10^{6}$ & $113\times113$ & 4.67 & 3.68 & 6.62 & 4.66 & $5\times10^{8}$ & $561\times561$ & 27.01 & 13.46 & 149.78 & 62.75 \\
$2\times10^{6}$ & $127\times127$ & 5.51 & 4.26 & 9.70 & 6.25 & $1\times10^{9}$ & $561\times561$ & 33.33 & 15.79 & 222.84 & 86.60 \\
$5\times10^{6}$ & $127\times127$ & 6.72 & 5.14 & 15.38 & 9.16 & $2\times10^{9}$ & $561\times561$ & 41.01 & 18.43 & 337.69 & 116.62 \\
$1\times10^{7}$ & $141\times141$ & 7.76 & 5.93 & 15.84 & 12.18 & $5\times10^{9}$ & $561\times561$ & 53.74 & 23.82 & 561.72 & 184.19 \\
$2\times10^{7}$ & $169\times169$ & 9.28 & 6.82 & 18.58 & 16.13 & $1\times10^{10}$ & $841\times841$ & 66.17 & 27.63 & 822.24 & 245.95 \\
$5\times10^{7}$ & $169\times169$ & 12.00 & 8.20 & 42.25 & 23.33 & $2\times10^{10}$ & $841\times841$ & 81.00 & 32.03 & 1268.84 & 328.35 \\
$1\times10^{8}$ & $211\times211$ & 15.62 & 9.42 & 57.62 & 30.78 & $5\times10^{10}$ & $841\times841$ & 106.32 & 40.21 & 1232.70 & 486.81 \\
$2\times10^{8}$ & $561\times561$ & 19.84 & 10.91 & 83.90 & 41.14 &  &  &  &  &  &  \\

			\bottomrule
		\end{tabular}%
	}
\end{table*}

\begin{table*}[!htbp]
	\centering
	\caption{Simulation parameters, grid resolutions, Nusselt numbers ($Nu$), and Reynolds numbers ($Re$). The Rayleigh number is fixed at $Ra = 1\times10^8$.}
	\label{tab:Pr_dependence}

	\footnotesize
	\setlength{\tabcolsep}{3.5pt}
	\renewcommand{\arraystretch}{1.15}

	\resizebox{\textwidth}{!}{%
		\begin{tabular}{@{}c c c c c c @{\hspace{1.2em}} c c c c c c@{}}
			\toprule
			$Pr$ & $N_x \times N_z$
			& \multicolumn{2}{c}{$Nu$}
			& \multicolumn{2}{c}{$Re$}
			&
			$Pr$ & $N_x \times N_z$
			& \multicolumn{2}{c}{$Nu$}
			& \multicolumn{2}{c}{$Re$} \\
			\cmidrule(lr){3-4}
			\cmidrule(lr){5-6}
			\cmidrule(lr){9-10}
			\cmidrule(lr){11-12}
			&
			& EXT-NE & MON-NE
			& EXT-NE & MON-NE
			&
			&
			& EXT-NE & MON-NE
			& EXT-NE & MON-NE \\
			\midrule
			0.03125 & $701\times701$ & 9.35  & --   & 16198.02 & --      & 1  & $211\times211$ & 14.10 & 8.98 & 732.30 & 322.53 \\
			0.03125 & $561\times561$ & 9.38  & 6.20 & 16032.80 & 5838.16 & 2  & $211\times211$ & 14.80 & 9.18 & 379.97 & 168.01 \\
			0.0625  & $561\times561$ & 10.42 & 6.95 & 9821.98  & 3455.19 & 4  & $211\times211$ & 15.47 & 9.31 & 191.64 & 86.68  \\
			0.125   & $561\times561$ & 11.61 & 7.72 & 6432.82  & 2056.47 & 8  & $211\times211$ & 15.56 & 9.39 & 88.29  & 44.19  \\
			0.25    & $561\times561$ & 12.60 & 8.29 & 3136.96  & 1159.51 & 16 & $211\times211$ & 15.46 & 9.45 & 38.23  & 22.34  \\
			0.5     & $561\times561$ & 13.37 & 8.68 & 1361.06  & 617.84  & 32 & $211\times211$ & 15.07 & 9.48 & 18.06  & 11.25  \\
			\bottomrule
		\end{tabular}%
	}
\end{table*}

\clearpage
\bibliography{References}

\end{document}